\documentclass[traditabstract,a4]{aa}  

\usepackage{epsfig}
\usepackage{graphicx}
\usepackage{epstopdf}
\usepackage{color}  
\usepackage{array}   
\usepackage{units}  
\usepackage[breaklinks, colorlinks, citecolor=blue]{hyperref}
\usepackage{natbib}  
\usepackage{multirow}   
\usepackage{amsmath,amssymb}  
\usepackage[english]{babel}
\usepackage[latin1]{inputenc}
\usepackage[T1]{fontenc}
\usepackage{longtable,lscape}
\usepackage{footnote}
\usepackage{txfonts}
\usepackage[toc,page]{appendix} 

\begin{document}

\newcommand{\avg}[1]{\left< #1 \right>}


   \title{Mapping the temperature\\ of the intra-cluster medium with the tSZ effect.} 

   \author{G. Hurier\inst{1} \& C. Tchernin\inst{2}
          }

\institute{Centro de Estudios de F\'isica del Cosmos de Arag\'on (CEFCA),Plaza de San Juan, 1, planta 2, E-44001, Teruel, Spain
\and Center  for  Astronomy,  Institute  for  Theoretical  Astrophysics,  Heidelberg  University,  Philosophenweg  12,  69120  Heidelberg,
Germany\\
\\
\email{ghurier@cefca.es} 
}

   \date{Received /Accepted}
 
   \abstract{The hot electrons in the intra-cluster medium produce a spectral distorsion of the cosmic microwave background (CMB) black body emission, the thermal Sunyaev-Zel'dovich effect (tSZ). This characteristic spectral distorsion is now commonly used to detect and characterize the properties of galaxy clusters. The tSZ effect spectral distorsion does not depends on the redshift, and is only slightly affected by the galaxy cluster properties via the relativistic corrections, when the electrons reach relativistic velocities. The present work proposes a linear component separation approach to extract the tSZ effect Compton parameter and relativistic corrections for next-generation CMB experiments. We demonstrated that relativistic corrections, if neglected, would induce a significant bias on galaxy cluster Compton parameter, tSZ scaling relation slope, and tSZ angular power spectrum shape measurements. We showed that tSZ relativistic corrections mapping can be achieved at high signal-to-noise ratio with a low level of contamination up to $\ell=3000$ for next generation CMB experiments. At smaller angular scales the contamination produced by infra-red emission will be a significant source of bias. Such tSZ relativistic corrections mapping enables the study of galaxy cluster temperature profile via the tSZ effect only.}

  \maketitle


\section{Introduction}

The thermal Sunyaev-Zel'dovich (tSZ) \citep{sun72} is produced by the interaction of cosmic microwave background (CMB) photons with the hot ionized plasma inside galaxy clusters. This effect induces a characteristic spectral distorsion of the CMB black body spectral energy distribution (SED).\\
Since the last five years, the measurement of the tSZ effect produced by galaxy clusters have reached an unprecedented accuracy \citep{mar11,ble15,psz2}. 
It is generally assumed that the tSZ spectral distorsion is invariant with the galaxy cluster redshift and intra-cluster medium (ICM) physical properties. The redshift invariance of the tSZ effect is satisfied for adiabatic evolution of the universe and has been confirmed by recent measurement \citep[see e.g.,][]{hur14}.
However, the invariance with respect to the plasma properties is broken if the electrons of the hot plasma reach relativistic velocities, and relativistic corrections have to be applied \citep{wri79}. These relativistic corrections become significant for temperature above a few keV (electron velocities of $\simeq 0.1\, c$ at $5$ keV) and enable the possibility to use the tSZ effect spectral distorsion as a probe to measure the temperature of the hot plasma inside galaxy clusters \citep{poi98,ens04}.\\

The extraction of the tSZ effect is a challenge, the intensity of the effect on the sky is small, around $10^{-6}-10^{-4} \, K_{\rm CMB}$, and present strong correlation with other astrophysical processes, such as radio-loud AGN emission, infra-red emission from galaxy cluster member galaxies \citep{planckszcib}.
Nevertheless, the construction of tSZ effect Compton parameter maps through component separation is now a mature activity \citep{hur13,rem11,bob08}, and reach an angular resolution of 1.4 arcmin ({\color{blue} Aghanim et al. 2016 in prep}) combining Planck and ACT data. 
However, the relativistic corrections to the tSZ SED are one order of magnitude smaller than the tSZ effect itself and can be easily confused with other astrophysical processes for experiments that have an unsufficient spectral coverage in the range $[10,1000]\, {\rm GHz}$.\\

For instance, evidence of the tSZ relativistic corrections have been claimed with Z-Spec \citep{zem10,zem12} and 
present experiment such as Planck \citep{planckmis} have enabled the statistical detection at $5 \sigma$ of the tSZ effect relativistic corrections ({\color{blue} Hurier 2016 accepted by A\&A}). The future generation of high resolution high sensitivity experiments such as COrE+\footnote{\url{http://hdl.handle.net/11299/169642}} will strongly improve the sensitivity and the number of observed frequencies, opening the door to tight measurements of the tSZ relativistic corrections and scientific exploitation of these relativistic corrections.\\

This paper is dedicated to the reconstruction and mapping of the tSZ relativistic corrections with the MILCA component separation method \citep{hur13}, previously used to reconstruct fullsky maps of the tSZ effect \citep{planck13szmap,planck15szmap}.
The scientific outcomes of the relativistic corrections mapping will be discussed in an upcoming paper ({\color{blue} Tchernin \& Hurier in prep.}).\\
The paper is organized as follows: Sect.~\ref{skysim} describes the tSZ effects and the relativistic corrections to this effect. In Sect.~\ref{methodo} we describe the tSZ relativistic corrections maps reconstruction methodology and characterize the transfert function of the reconstructed maps. Then Sect.~\ref{secbias} adresses the drawbacks of neglecting tSZ relativistic corrections for future CMB experiments. Finally, we discuss the main limitation of the tSZ relativistic corrections recovery in Sect.~\ref{secsyst} and we draw our conclusions in Sect.~\ref{concl}

\section{The tSZ effect}
\label{skysim}
\label{sze}
Through the thermal Sunyaev-Zel'dovich effect \citep{sun72} CMB photons receive an average energy boost by collision with hot (a few keV) ionized electrons of the intra-cluster medium \citep[see e.g.][for reviews]{bir99,car02}.
The intensity of this effect toward a given line of sight is given by the thermal SZ Compton parameter,
\begin{equation}
y  = \int n_{e} \frac{k_{\rm{B}} T_{\rm{e}}}{m_{\rm{e}} c^{2} } \sigma_{\rm T} \  \rm{d}{\it l}
\label{comppar}
\end{equation}
where d${\it l}$ is the distance along the line of sight, and $n_{\rm{e}}$
and $T_{e}$ are the electron number density and temperature,
respectively.
In units of CMB temperature the contribution of the tSZ effect for a given observation frequency $\nu$ is
\begin{equation}
\frac{\Delta T_{\rm{CMB}}}{T_{\rm{CMB}} }= g_\nu \ y.
\end{equation}
Neglecting relativistic corrections we have 
\begin{equation}
g_{\nu} \simeq \left[ x\coth \left(\frac{x}{2}\right) - 4 \right],
\label{szspec}
\end{equation}
with $ x=h \nu/(k_{\rm{B}} T_{\rm{CMB}})$. At $z=0$, where $T_{\rm CMB}(z=0)$~=~2.726$\pm$0.001~K, the tSZ effect is negative below 217~GHz and positive for higher frequencies. This characteristic spectral signature of tSZ makes it a unique tool for the detection of galaxy clusters.\\

This spectral signature is also slightly dependent of $T_{\rm e}$ through relativistic corrections \citep[see][for a detailed fitting formula]{ito00}.
In the present work, the relativistic corrections on the tSZ emission law have been computed as presented in \citet{poi98}.
Based on this estimation, we assume that the relativistic correction on the tSZ emission law can be described as a first order linear approximation\footnote{We note that this assumption is satisfied if the tSZ spectral distorsion variations are monotonic with respect to $T_{\rm e}$.},
\begin{equation}
\Delta T^{\rm relat}_{\rm CMB}(T_e) \simeq \Delta T^{\rm unrelat}_{\rm CMB} + T_e \Delta T^{\rm cor}_{\rm CMB},
\end{equation}
where $\Delta T^{\rm relat}_{\rm CMB}(T_e)$ is the total tSZ induced CMB distorsion, $\Delta T^{\rm unrelat}_{\rm CMB}$ is the non-relativistic tSZ effect contribution, and $\Delta T^{\rm cor}_{\rm CMB}$ is the relativistic correction per units of electronic temperature, $T_e$, assuming a first order linear approximation.
Given that, the averaged tSZ emission from the electron populations at various temperatures can be modeled as a single temperature.\\
This assumption is already implicitly applied when fitting a single temperature to tSZ effect observed on the sky. Indeed, due to the line of sight integration of the tSZ signal, the observed emission is already a linear combination of tSZ effect produced by electrons at different electronic temperatures.

\begin{figure}[!th]
\center
\includegraphics[width=9cm]{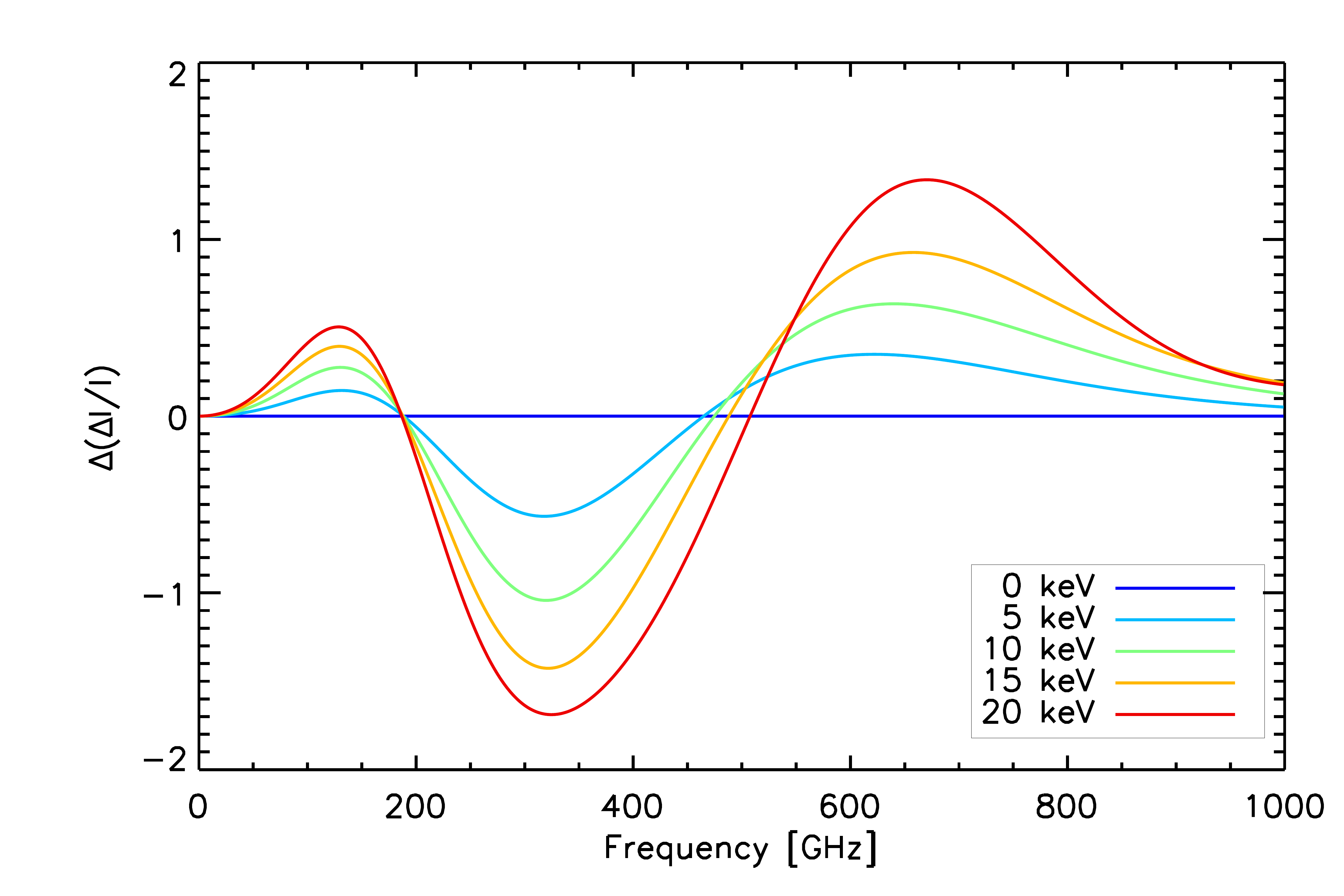}
\includegraphics[width=9cm]{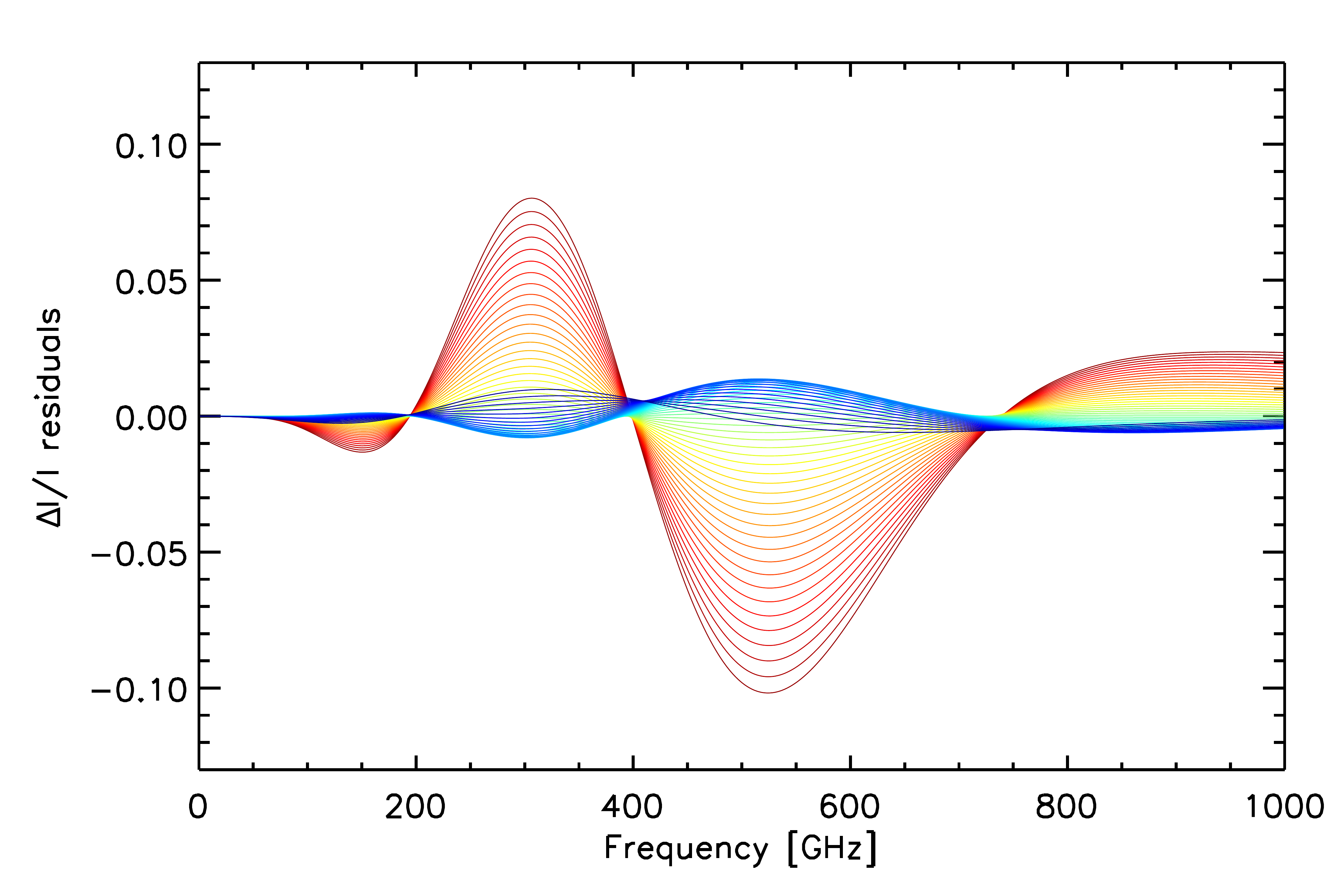}
\caption{Top panel: tSZ relativistic corrections to the tSZ spectral distorsion as a function of the frequency for various temperature of the hot plasma from 0 to 20 keV. Bottom panel: error on the tSZ spectral distorsion induced by the linear approximation from 0 (dark blue solid line) to 10 keV (dark red solid line) as a function of the frequency, using $T_e = 5$ keV as reference temperature.}
\label{szr}
\end{figure}

Figure~\ref{szr} (top panel) shows the relativistic corrections additional spectral distorsion to the non-relativistic tSZ spectral distorsion as a function of the frequency for various temperatures of the hot plasma ranging from 0 to 20 keV. 
Contrary to the non-relativistic tSZ effect, tSZ relativistic corrections are not null at $\nu \simeq 217$ Ghz, but at $\simeq 188$ GHz. Consequently, tSZ relativistic corrections induce a shift of the tSZ effect zero frequency, $\nu_0$, that can be parametrized as $\nu_0 \simeq 217.4 + T_e/2$ GHz with $T_e$ expressed in keV. 
The relativistic corrections essentially reduce the amplitude of the spectral distorsion for frequencies up to $\simeq 450$ GHz and increase the tSZ distorsion for higher frequencies. Considering that tSZ effect is usually estimated from low frequency channels (below 500 GHz, where CMB emission is significant compared to other emissions on the sky), the tSZ relativistic corrections might induce a bias on measured Compton parameter that is under-estimated if relativistic corrections are neglected.\\
We also present on Fig.~\ref{szr} (bottom panel) the error on the tSZ spectral distorsion that is induced by our linear approximation.
We show that the error on the tSZ spectral distorsion is more than one order of magnitude below than the amplitude of the relativistic correction itself. It demonstrates that the linear approximation provides a good description of the tSZ spectral distorsion from 0 to 10 keV, that is the typical temperature range of galaxy clusters. We note that in a high temperature situation, the reference temperature $T_e$ can be changed to reduce the discrepancy for high temperature intra-cluster medium.

\section{Mapping the tSZ relativistic corrections}
\label{methodo}

 To map the tSZ effect relativistic corrections, we used a component separation approach based on variance minimization \citep{hur13}. This methods has been successfully used to recover the tSZ effect Compton parameter with Planck data \citep{planckszs} and with combined ACT and Planck data ({\color{blue} Aghanim et al. in prep}).
 In this section, we describe the hypothesis made and methodology used to reconstruct the tSZ effect relativistic corrections, and we carefully study the key frequencies for foreground/background emissions removal.
 
\subsection{MILCA approach}

We assumed that the sky can be well described as a linear mixture of astrophysical sources,
\begin{align}
{\bf T} = {\cal A} {\bf S} + {\bf N},
\end{align}
Where ${\bf T}$ are the observed frequency channels, ${\cal A}$ is the mixing matrix, ${\bf S}$ are the astrophysical components and ${\bf N}$ is the instrumental noise.
The MILCA method is, by construction, a linear methods aiming at extracting a given component, knowing its spectral behavior, from multi-channel observations of the sky. However, the relativistic corrections to the tSZ effect are not linear.\\ 
Consequently, to apply the MILCA approach on the tSZ relativistic corrections reconstruction, we first performed a linear approximation of tSZ relativistic corrections SED. We modelled the tSZ spectral distorsion, $g_{\nu,T_e}$, as follows
\begin{align}
g_{\nu,T_e} \simeq g_{\nu,0} + T_e h_{\nu},
\end{align}
Where $g_{\nu,T_e}$ is the tSZ spectral distorsion presented in sect.~\ref{sze} at the temperature, $T_e$, of the hot electrons in the intra-cluster medium, and $h_\nu$ is a linear approximation of the tSZ relativistic corrections.\\
In the present analysis we computed $h_\nu$ as
\begin{align}
\label{linap}
h_\nu = \frac{g_{\nu,T_1} - g_{\nu,0}}{T_1},
\end{align}
for $T_1 = 5$ keV, considering that the typical temperature of galaxy clusters is a few keV. To increase the accuracy of the reconstruction method $T_1$ can be set to any temperature depending on the concerned galaxy cluster average temperature.\\

\begin{figure*}[!th]
\begin{center}
\includegraphics[scale=0.35]{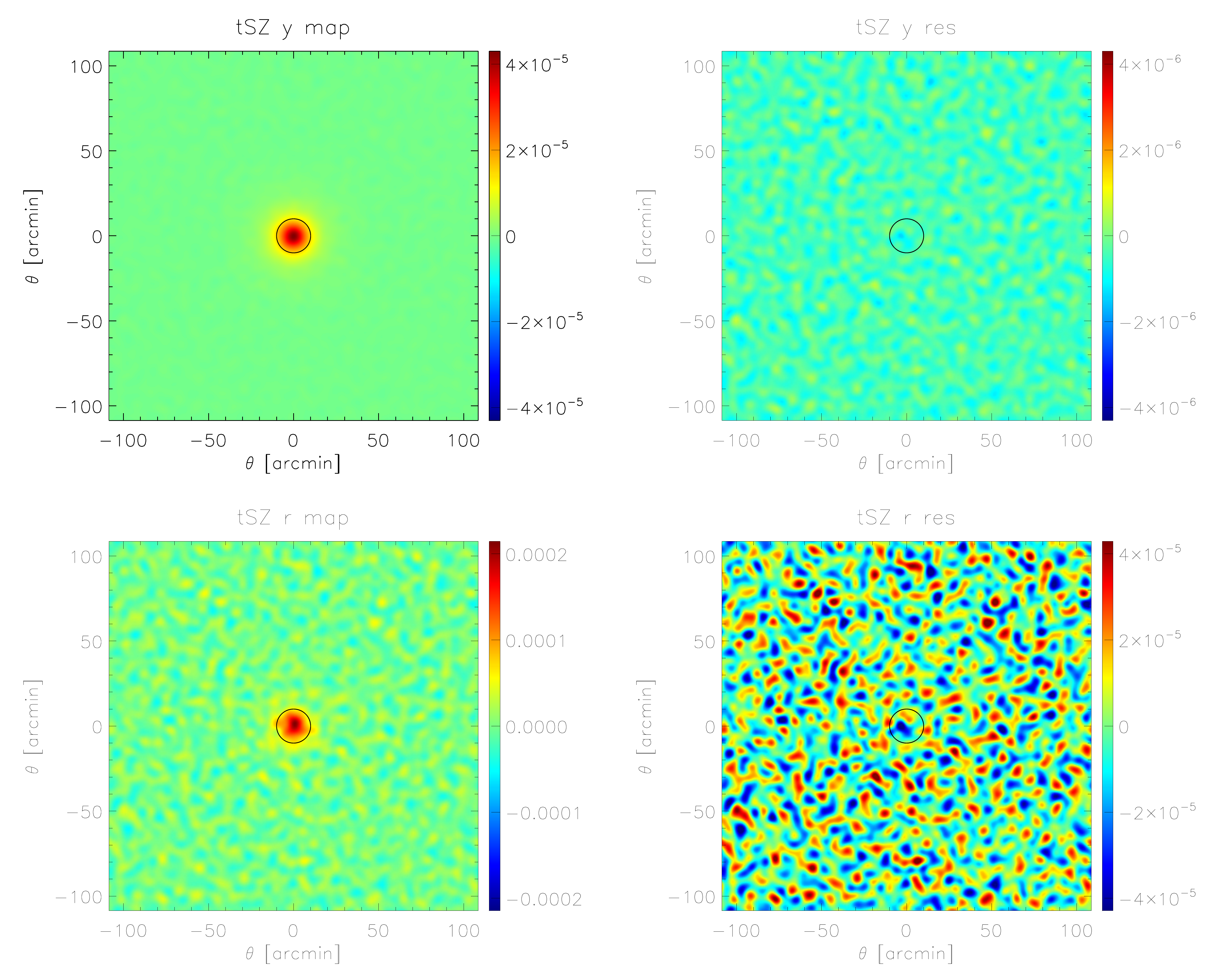}
\caption{From left to right and top to bottom: reconstructed tSZ $y$-map, residuals in the reconstructed $y$-map, reconstructed tSZ $r$-map, residuals in the reconstructed $r$-map for a patch of $3.5\times3.5$ degrees, with 0.4 arcmin pixels, centered on a simulated galaxy clusters at 7 arcmin FWHM angular resolution}.
\label{simures}
\end{center}
\end{figure*}

The MILCA method achieve component separation through a minimization of the variance of the reconstructed map under constraints. It allows to perform a localized reconstruction of the astrophysical component both in real and Fourrier spaces.\\
The most simple estimator of tSZ relativistic corrections reads
\begin{align}
S_{r} = \left({\bf f}^{\rm T}_\nu {\cal C}^{-1}_{\bf T} {\bf f}_\nu \right)^{-1} {\bf f}^{\rm T}_\nu {\cal C}^{-1}_{\bf T} {\bf T},
\end {align}
with $S_{r}$ the relativistic corrections map, ${\bf f}^{\rm T}_\nu$ it's frequency dependence, and ${\cal C}^{-1}_{\bf T}$ the covariance matrix of the frequency channel maps ${\bf T}$. This estimator is a standard linear fit that minimizes the variance of the reconstructed signal.\\
However, it is well known that variance minimisation is biased by spatial correlations between the component to extract and other astrophysical components \citep{hur14}.
In the present case, the Compton parameter, $y$, and the relativistic corrections signal $r$ are, by construction, highly spatially correlated. Consequently, we need to avoid bias on the $r$-map by adding an extra constraints\footnote{Extra-constraints can be added for any astrophysical component that presents a well known spectral behavior, which is the case, for example of non-relativistic tSZ effect and CMB emission.} that remove the contribution from non-relativistic tSZ effect.
The estimator becomes
\begin{align}
{\bf S} = \left({\cal F}^{\rm T} {\cal C}^{-1}_{\bf T} {\cal F} \right)^{-1} {\cal F}^{\rm T}_\nu {\cal C}^{-1}_{\bf T} {\bf T},
\end {align}
where ${\cal F}$ is a rectangular matrix that contains the spectral behavior of the $n_c$ components on which a constraint is applied, and ${\bf S}$ is a vector that contains the $n_c$ maps corresponding to the constrained astrophysical components. We note that $S_{r} = {\bf e}^{\rm T}_{r} {\bf S}$, with ${\bf e}_{r}$ a vector that selects the subspace corresponding to the tSZ relativistic corrections component.\\
Additionally, the frequency maps covariance matrix, ${\cal C}_{\bf T}$, is not known and have to be estimated from the data. This matrix can be decomposed as, 
\begin{align}
{\cal C}_{\bf T} = {\cal A}^{\rm T} {\cal C}_{\bf S} {\cal A} + {\cal C}_{\bf N},
\end{align}
the astrophysical component covariance matrix ${\cal C}_{\bf S}$ and the instrumental noise covariance matrix ${\cal C}_{\bf N}$ do not have the same eigenvectors. Consequently, ${\cal C}_{\bf N}$ adds non-physical correlations to ${\cal C}_{\bf S}$. Similarly to physical correlations, these correlations will add some bias to the variance minimisation process. Consequently, we apply the following transformation,
\begin{align}
{\tilde {\cal C}}_{\bf T} = {\cal C}_{\bf T} - {\cal C}_{\bf N} - {\cal F} \left( {\cal F}^{\rm T} {\cal C}^{-1}_{\bf T} {\cal F} \right)^{-1} {\cal F}^{\rm T},
\label{ortho}
\end{align}
to remove noise-induced-bias and suppress the contribution from constrained components to the covariance matrix.
Then, we performed a rank reduction on ${\tilde {\cal C}}_{\bf T}$, keeping only the highest eigenvalues corresponding to the number of unconstrained astrophysical components to minimize.
Other eigenvalues are constrained by minimizing the noise contribution in the reconstructed map
\begin{align}
V_{{\bf N},r} = {\bf e}^{\rm T}_{r} {\cal W_{\rm r}}^{\rm T} {\cal C}_{\bf N} {\cal W_{\rm r}} {\bf e}_{r},
\end{align}
with ${\cal W_{\rm r}}^{\rm T} = \left({\cal F}^{\rm T} {\tilde {\cal C}}^{-1}_{\bf T} {\cal F} \right)^{-1} {\cal F}^{\rm T}_\nu {\tilde {\cal C}}^{-1}_{\bf T} $. We remind that the minimization is performed on the eigenvalues suppressed by the rank reduction of  ${\tilde {\cal C}}_{\bf T}$.\\
Finally the tSZ relativistic corrections maps is given by
\begin{align}
S_{\rm r} =  {\bf e}^{\rm T}_{r} {\cal W_{\rm r}}^{\rm T} {\bf T}.
\end {align}
Similarly, the Compton parameter map is obtained by minimizing
\begin{align}
V_{{\bf N},y} = {\bf e}^{\rm T}_{y} {\cal W_{\rm y}}^{\rm T} {\cal C}_{\bf N} {\cal W_{\rm y}} {\bf e}_{y},
\end{align}
with ${\bf e}_{y}$ a vector that selects the subspace corresponding to the tSZ Compton parameter map.
The $y$-map is computed as
\begin{align}
S_{\rm y} =  {\bf e}^{\rm T}_{y} {\cal W_{\rm y}}^{\rm T} {\bf T}.
\end {align}
This approach allows to build a tSZ Compton parameter map $y  = \frac{k_{\rm B} \sigma_{\rm T}}{m_e c^2} \int n_e T_e dl$, and 
a map of tSZ relativistic corrections $r  \simeq \frac{k_{\rm B} \sigma_{\rm T}}{m_e c^2} \int n_e T^2_e dl$, this dependency can be derived from Eq.~\ref{linap}. The tSZ distorsion at a given frequency is given by $\Delta T/T \simeq g_{\nu,0} y + h_\nu r$.\\

A more detailed description of the MILCA method can be found in \citet{hur13}.

\begin{figure*}[!th]
\begin{center}
\includegraphics[scale=0.15]{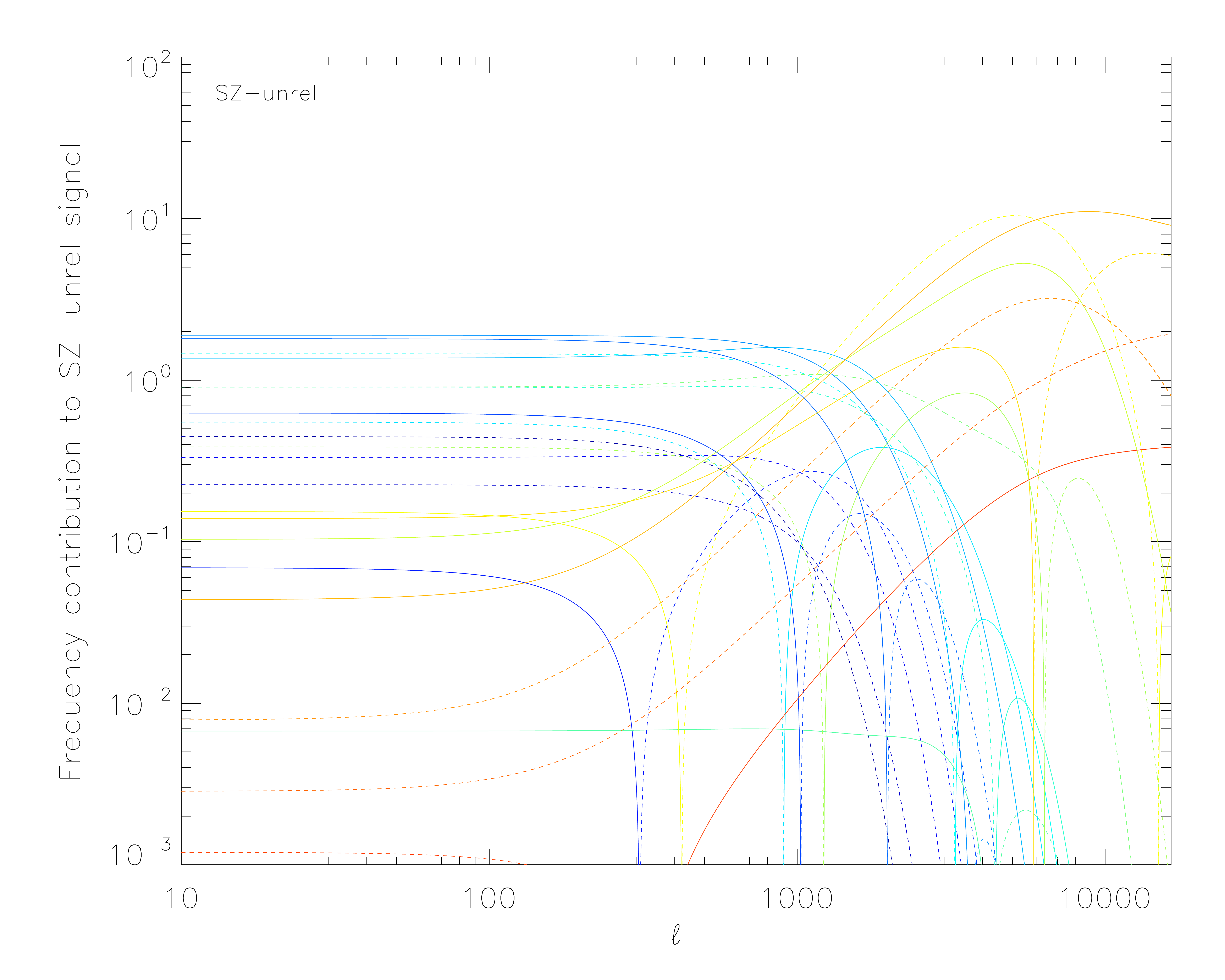}
\includegraphics[scale=0.15]{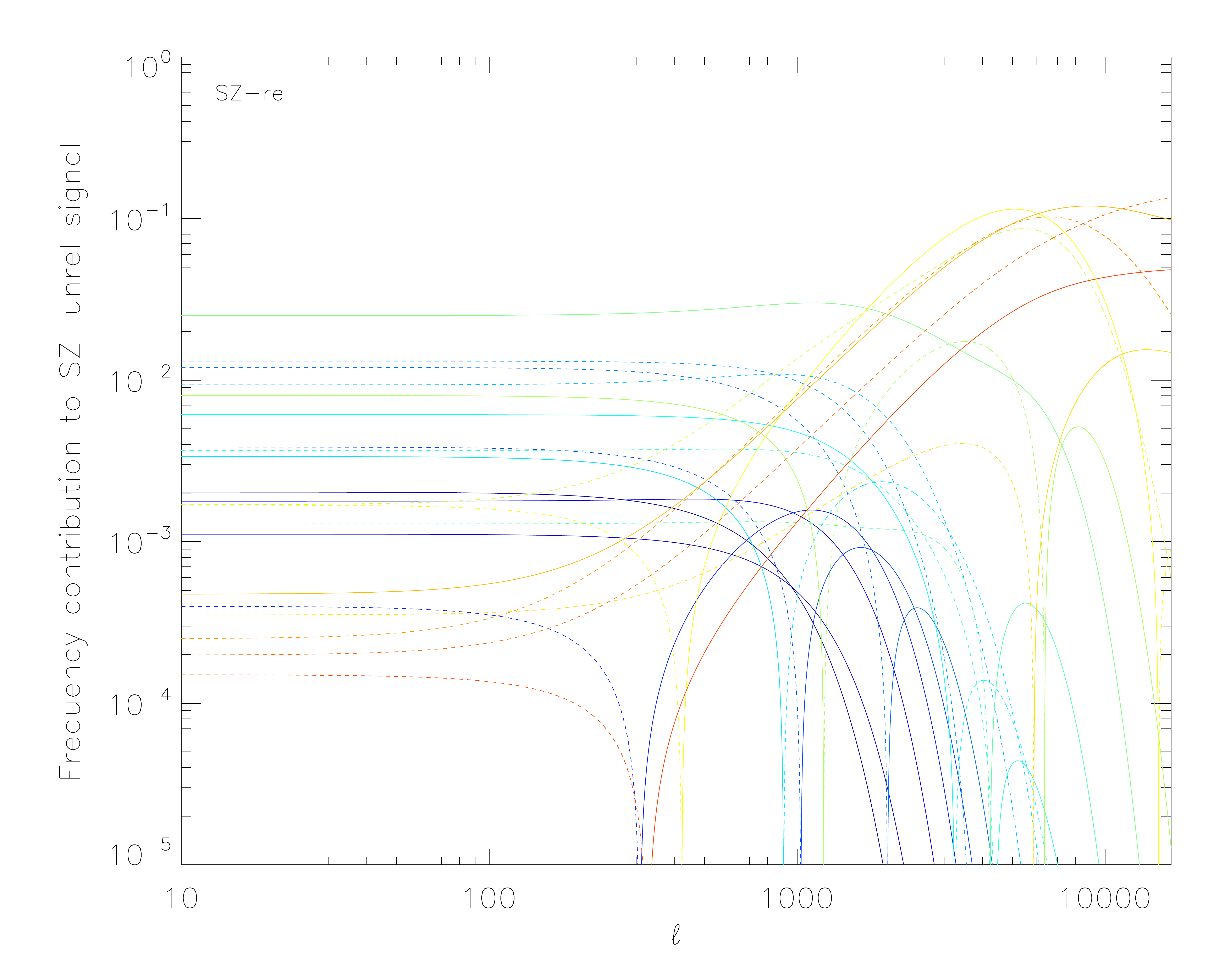}
\includegraphics[scale=0.15]{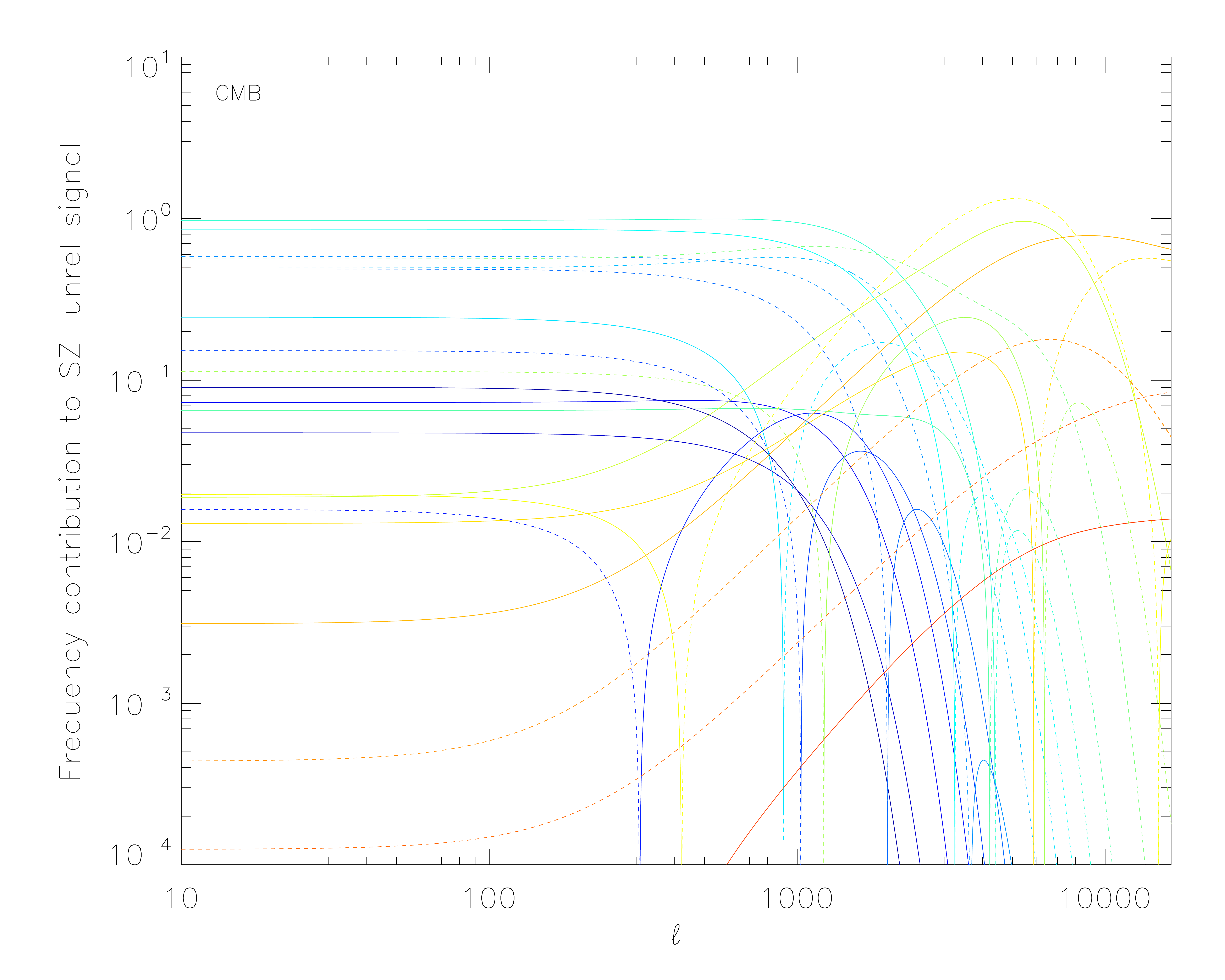}
\includegraphics[scale=0.15]{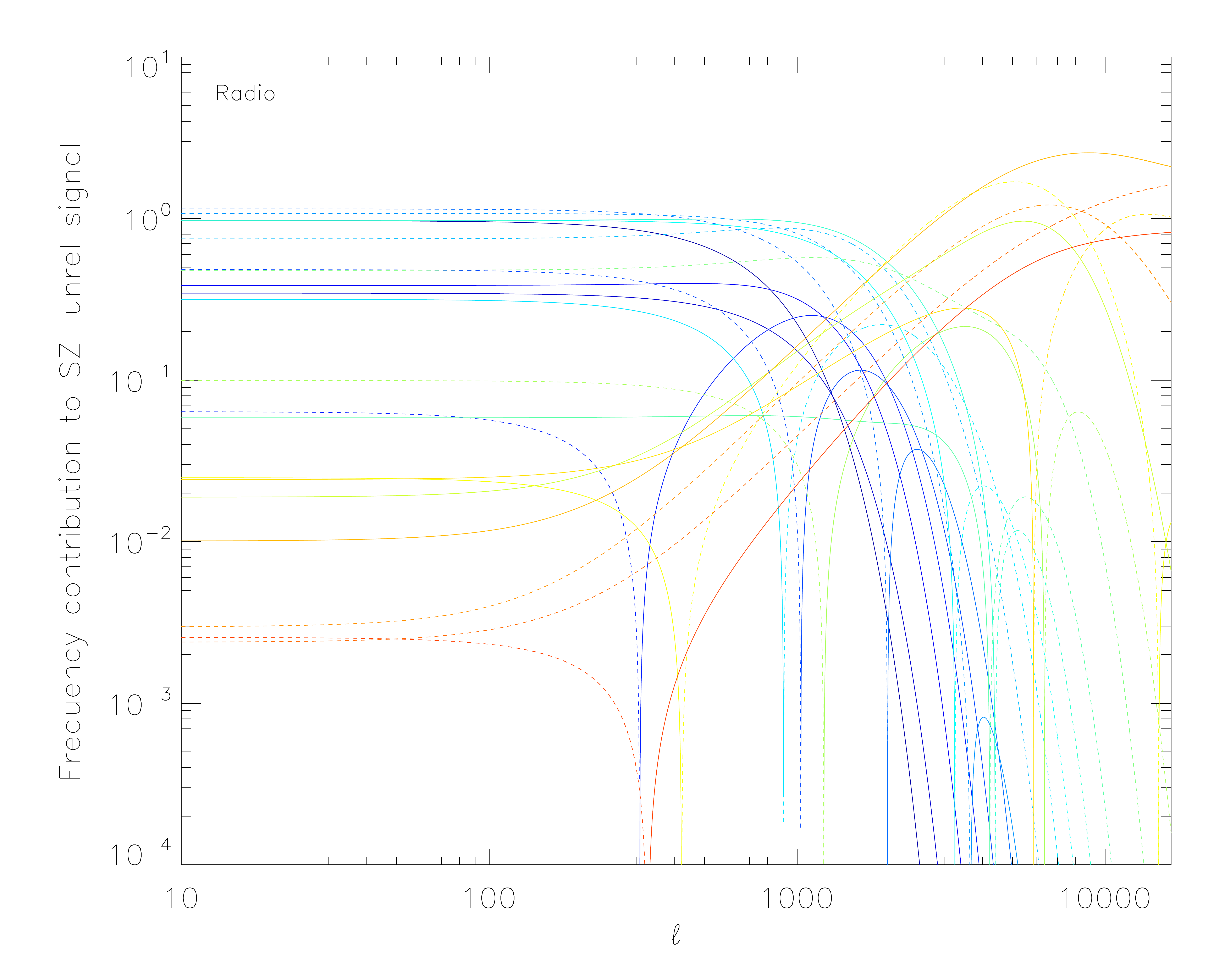}
\includegraphics[scale=0.15]{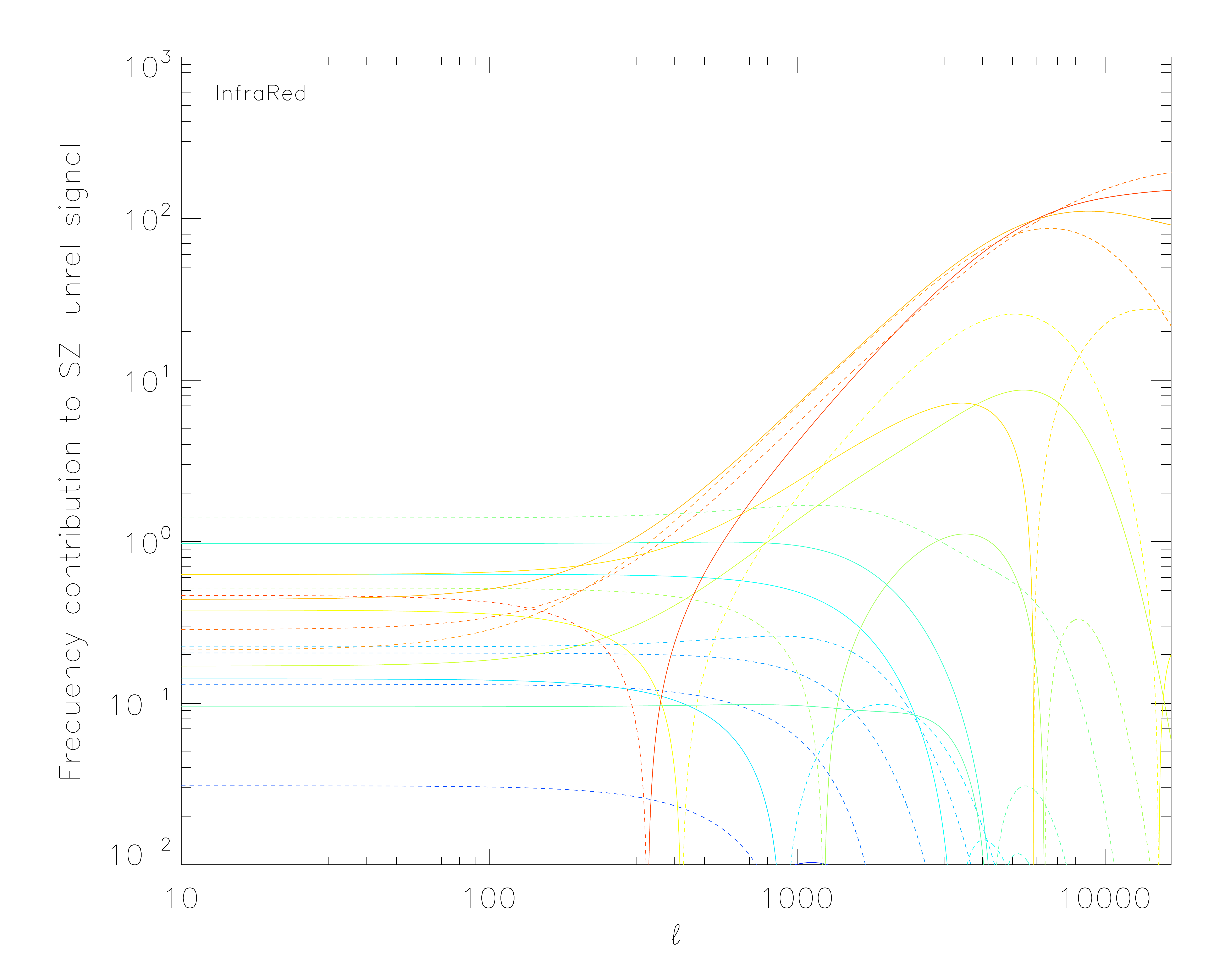}
\caption{Contributions received by the tSZ Compton parameter map as a function of the angular scale. From left to right and top to bottom: contribution of each frequency (from 30 GHz in dark blue to 800 GHz in red for the 21 frequency of COrE+) to non-relativistic tSZ effect, relativistic corrections to tSZ effect, CMB, Radio component, and Infrared components in the reconstructed map. Solid lines indicates positive contributions and dashed line negative contributions.}.
\label{winunrel}
\end{center}
\end{figure*}

\subsection{Test on simulated datasets}

\label{sims}
We are now interested in testing the algorithm described above.
Thus, we have constructed simulations of the full microwave and submillimeter sky as it will be observed by next generation CMB experiments like COrE+ (a summary of the planned caracteristic can be found in table.~\ref{tabc}) using a set of template maps to reproduce astrophysical components. A complete description of the simulation procedure can be found in \citet{hur13}.
Within the considered components, we include CMB, diffuse Galactic emissions (synchrotron, free-free and thermal dust), Galactic CO emission, Galactic and extragalactic point sources. We also added the cosmic infra-red background contribution following the best fitting model from \citet{planckszcib}.\\
Each template is assumed to be the real sky signal and is convolved by the point spread function of each COrE+ channels, and the component SEDs are convolved by the corresponding bandpass assuming $\Delta \nu / \nu = 0.2$.
Finally, we added an homogeneous instrumental noise accordingly to the COrE+ detector sensitivity for the 21 observed frequencies.
We note that MILCA weights are constructed to be insensitive to the addition or the substraction of tSZ signal in the input maps, under the assumtion that the tSZ signal is not spatially correlated with other astrophysical components (Eq.~\ref{ortho}) consequently, the MILCA weights can be estimated without any tSZ emission on the sky.
We stress that these simulations assume a single SED for each component (except for the CIB). We will discuss in the following the impact of varying SED on the sky.

\begin{table}
\label{tabc}
\center
\caption{Main characteristics (frequencies, beams, number of detector $N_{\rm det}$, and sensitivity) planned for the COrE+ experiment, for each channel the bandwidth is $\Delta \nu / \nu = 0.25$. }
\begin{tabular}{|c|c|c|c|}
\hline
channel [GHz] & beam [arcmin] & $N_{\rm det}$ & $\Delta T$ [$\mu$K.arcmin] \\
\hline
\hline
60 & 14 & 28 & 9.8 \\
70 & 12 & 30 & 9.1 \\
80 & 10.5 & 64 & 6.1 \\
90 & 9.33 & 102 & 4.8 \\
100 & 8.4 & 120 & 4.3 \\
115 & 7.3 & 196 & 3.4 \\
130 & 6.46 & 264 & 3.0 \\
145 & 5.79 & 388 & 2.5 \\
160 & 5.25 & 534 & 2.3 \\
175 & 4.8 & 554 & 2.4 \\
195 & 4.31 & 600 & 2.5 \\
220 & 3.82 & 490 & 3.2 \\
255 & 3.29 & 486 & 4.1 \\
295 & 2.85 & 260 & 8.1 \\
340 & 2.45 & 200 & 14.6 \\
390 & 2.15 & 120 & 33.7 \\
450 & 1.87 & 120 & 71.4 \\
520 & 1.62 & 120 & 181.1 \\
600 & 1.4 & 120 & 551 \\
700 & 1.2 & 60 & 3293.5 \\
800 & 1.05 & 60 & 14499.8 \\
\hline
\end{tabular}
\end{table}

We simulated a galaxy clusters following a universal pressure profil \citep{arn10,planckppp}, $P_e$, and a polytropic relation between temperature, $T_e$, and density, $n_e$, with a polytropic index $\delta = 1.2$ such as $P_e = n_e T_e = n_e^\delta$.
We divided the clusters into cells of $0.05\, R_{500}$, we computed the tSZ effect produced by each cells individually accordingly to the cells densities and temperatures, and we apply the corresponding spectral distorsion to the COrE+ simulated frequency channels by projecting the galaxy cluster tSZ effect on the line-of-sight. We stress that, due to relativistic corrections, the tSZ effect cannot be described with a single spectral distorsion and a Compton parameter map.\\
Under the approximation we are doing for the MILCA-based component separation, we should recover:
\begin{itemize}
\item a map of the integrated pressure on the line of sight,
\item a map of the integrated product of pressure and temperature on the line of sight.
\end{itemize}

Figure~\ref{simures} presents the reconstructed tSZ $y$~and~$r$-maps at 7 arcmin FWHM resolution for a nearby massive galaxy cluster ($M_{500} = 8 \times 10^{14}$ M$_\odot$, $z = 0.06$). This reconstruction is presented as an exemple and a proof of concept, in the following we rely on analytical estimations of the noise and residuals in the MILCA tSZ maps.
At this resolution the noise level in MILCA maps is $2.74 \, 10^{-7}$ in the $y$-map and $1.53 \, 10^{-5} \, {\rm keV}$ in the $r$-map. For the considered galaxy cluster, the $y$-signal peaks at $y=7.1 \times 10^{-5}$. Thus, the noise level in the $r$-map can be converted to a noise level on the temperature at the peak $\Delta T_e = 0.2$ keV. However, we note that this uncertainty will strongly increase in the outskirt of the galaxy cluster due to a fainter $y$ signal.

This result demonstrates that a COrE+ like experiments will achieve a noise level in Compton parameter maps 10 times lower than the Planck experiment at the same angular scale\footnote{We verified that using the same simulations for Planck data, we recover a noise level consistent with the public Planck tSZ maps.}.\\
We observe that the signal in the $r$-map is less extended than the signal in the $y$-map. Indeed the $r$-map is sensitive to the product $T_e^2 n_e$ and our simulation assumes a polytropic profile. Thus, we expect to observe a more compact emission in the $r$-map than in the $y$-map.
We did not observe significant bias in the residuals of the $y$-map and the $r$-map, this indicates that systematic effect produced by our linear approximation are not significant compared to the statistical noise level.

\begin{figure*}[!th]
\begin{center}
\includegraphics[scale=0.15]{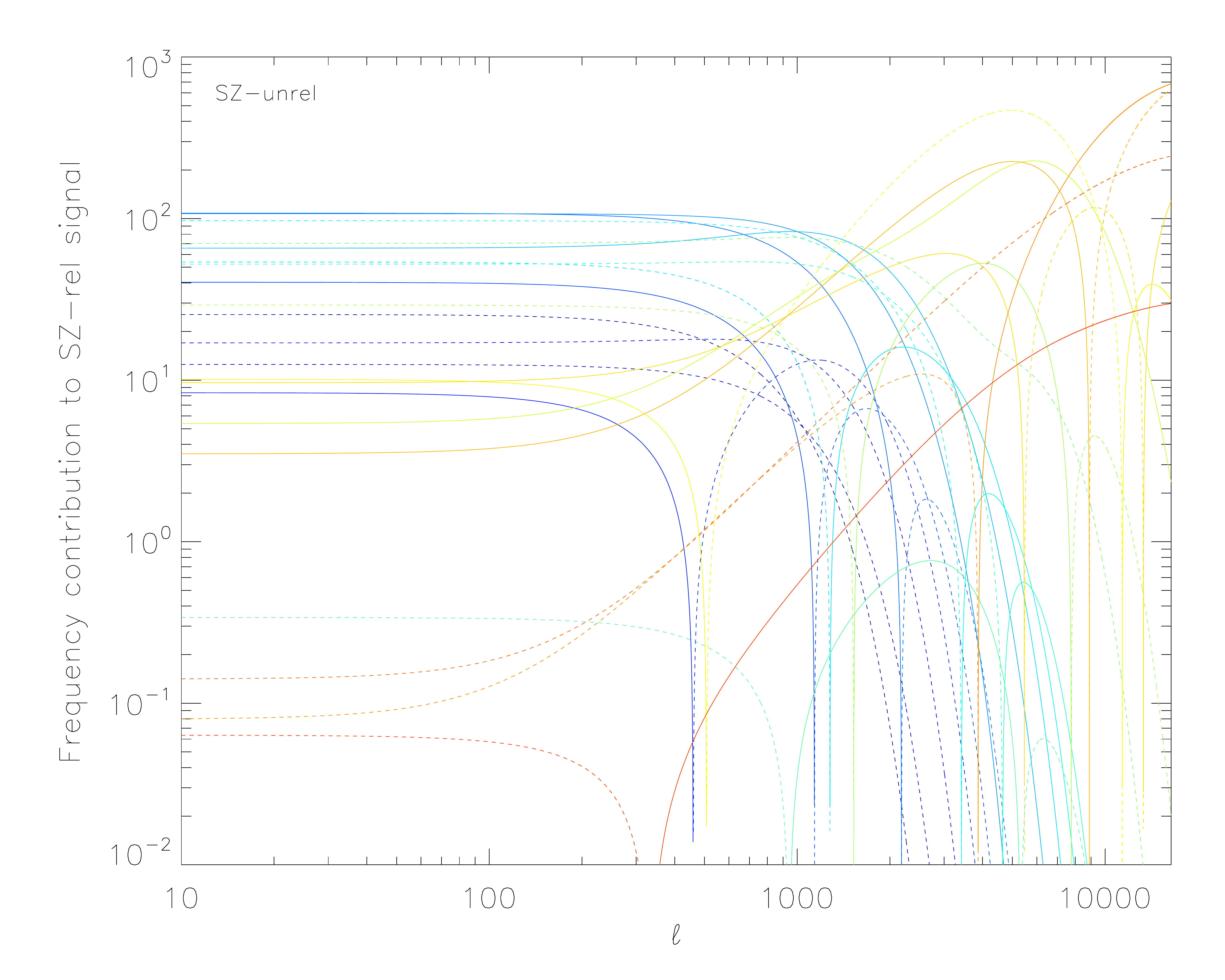}
\includegraphics[scale=0.15]{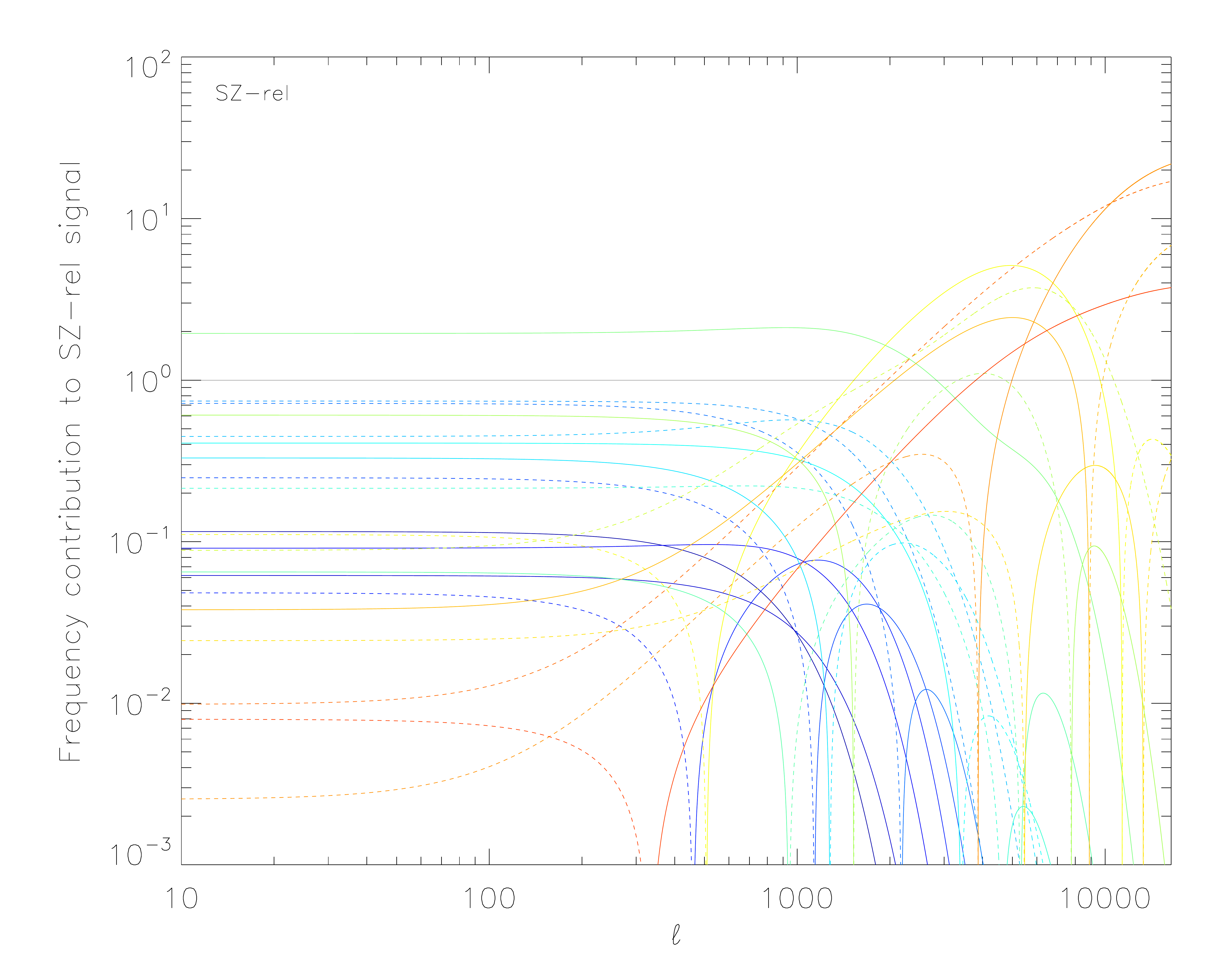}
\includegraphics[scale=0.15]{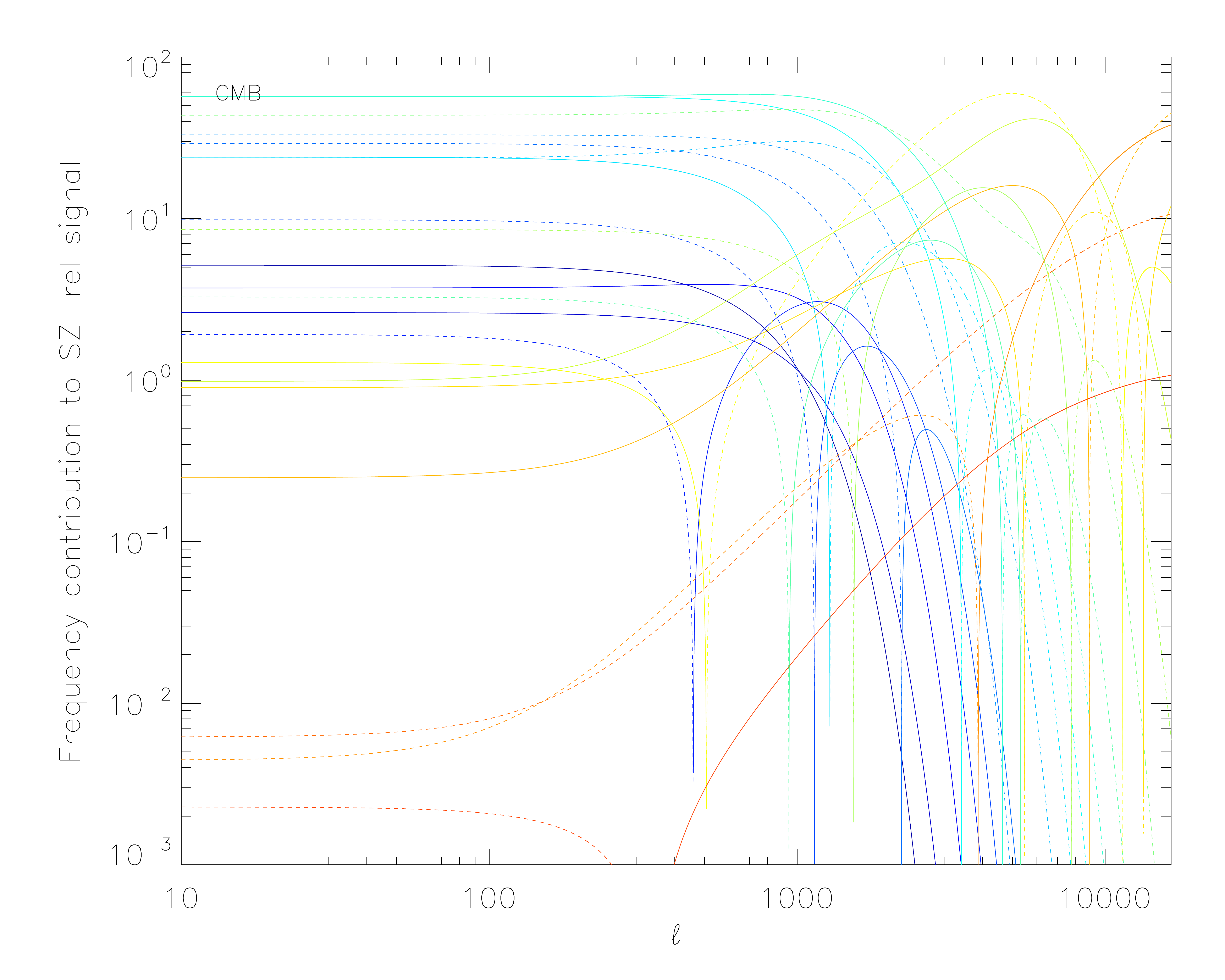}
\includegraphics[scale=0.15]{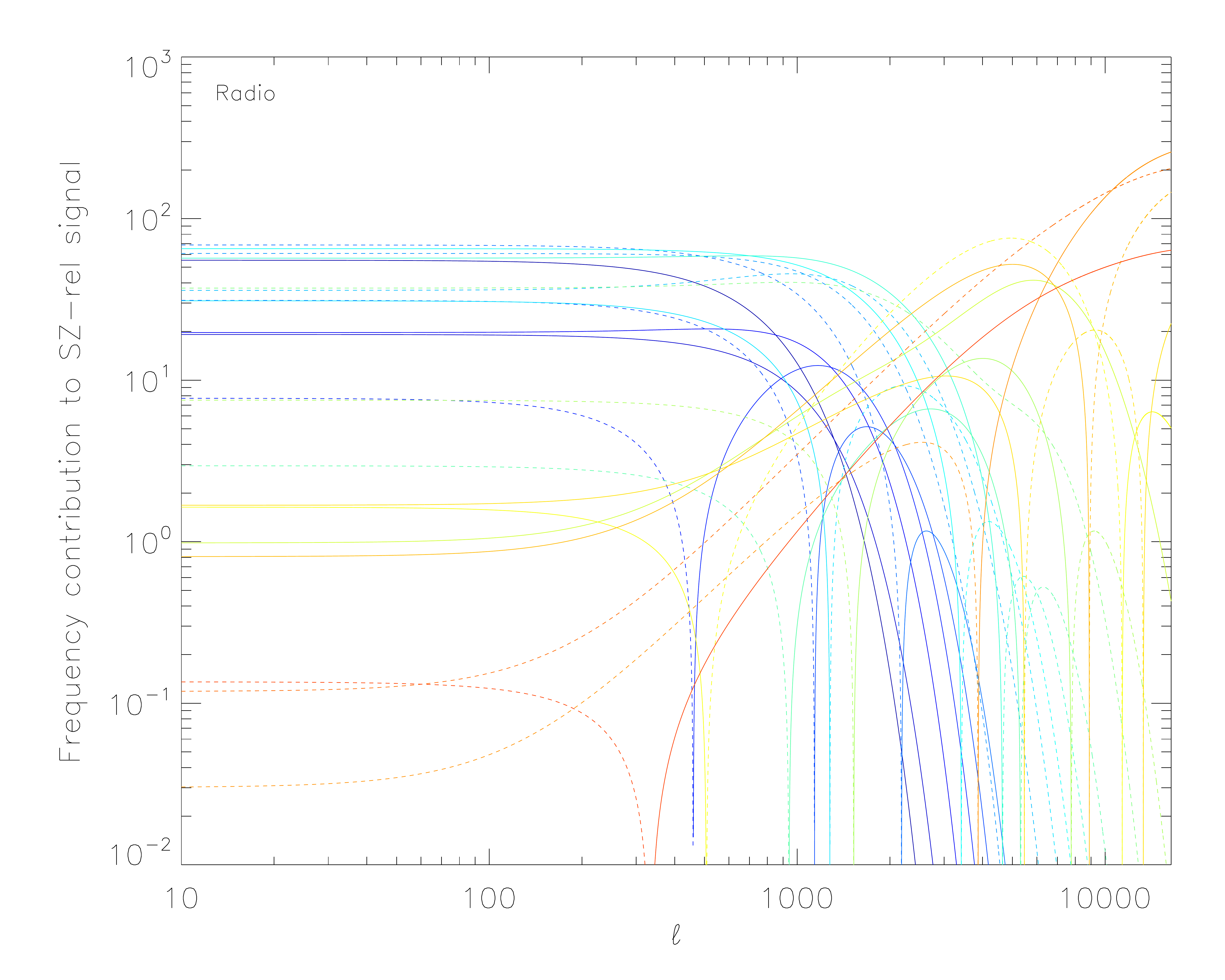}
\includegraphics[scale=0.15]{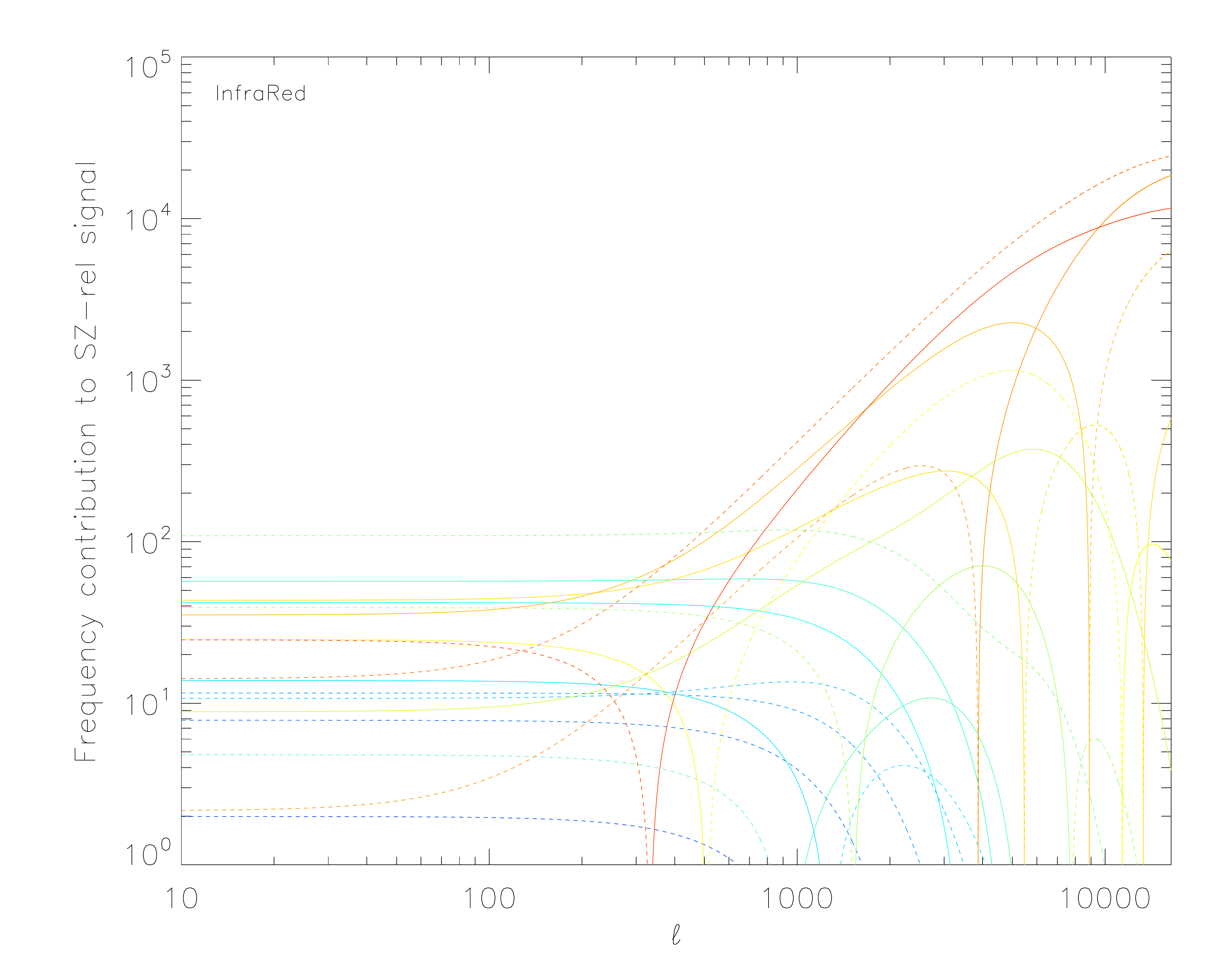}
\caption{Contributions received by the tSZ relativistic corrections map as a function of the angular scale. From left to right and top to bottom: contribution of each frequency (from 30 GHz in dark blue to 800 GHz in red for the 21 frequency of COrE+) to non-relativistic tSZ effect, relativistic correction to tSZ effect, CMB, Radio component, and Infrared components into the reconstructed map. Solid lines indicates positive contributions and dashed line negative contributions.}.
\label{winrel}
\end{center}
\end{figure*}

\subsection{Frequency contributions}

The MILCA method performs a localized component separation both in real and Fourrier spaces. The COrE+ setting involves different angular resolutions for the different frequency channels. Consequently, the optimal weights for the reconstruction intrinsically depend on the considered angular scale, $\ell$.\\
The transfert function of a given component, $c$, as a function of the scale, $\ell$, reads
\begin{align}
{\cal T}_{c}(\ell) =  {\bf e}^{\rm T}_{c} {\cal W}_{{\rm y}}^{\rm T}(\ell) {\cal F} {\bf e}_{c} .
\end{align}
Similarly, the contribution of the component, $i$, into the recovered map for the component $c$ reads
\begin{align}
\label{eqproj}
{\cal T}_{i,c}(\ell) =  {\bf e}^{\rm T}_{c} {\cal W}_{{\rm y}}^{\rm T}(\ell) {\bf F}_i ,
\end{align}
where ${\bf F}_i$ is the SED of the component $i$.
The contributions of the frequency $\nu$ to these transfert function are
\begin{align}
{\cal T}_{c}(\ell,\nu) =  \left[{\bf e}^{\rm T}_{c} {\cal W}_{{\rm y}}^{\rm T}(\ell){\bf e}_{\nu} \right] \left[ {\bf e}^{\rm T}_{\nu}  {\cal F} {\bf e}_{c}\right], \nonumber \\
{\cal T}_{i,c}(\ell,\nu) =  \left[{\bf e}^{\rm T}_{c} {\cal W}_{{\rm y}}^{\rm T}(\ell) {\bf e}_{\nu}\right] \left[ {\bf e}^{\rm T}_{\nu}{\bf F}_i\right],
\label{eqtrf}
\end{align}
with ${\bf e}_{\nu}$ a vector that selects the subspace corresponding to the frequency $\nu$. We note that ${\bf F}^T_i {\bf e}_{\nu}$ is a scalar quantity corresponding to the transmission of the component $i$ into the frequency channel $\nu$.\\
In order to determine the key frequency channels, we tracked the contributions of the 21 frequencies of COrE+ in the $y$-map and $r$-map reconstruction process.
{These contributions are estimated analytically using Eq.~\ref{eqtrf}, assuming a reference spectral behavior for each astrophysical components. The linear combination weights, ${\cal W}$, are computed from the fullsky simulations presented in Sect.~\ref{sims}}.

\subsubsection{MILCA tSZ $y$-map}

Figure~\ref{winunrel} presents the contribution of each frequency to the recovered tSZ Compton parameter map, we also present the contribution of each frequency to the suppression of tSZ relativistic corrections, CMB, Radio, and Infra-red components.\\
By construction, the sum of all the curves in the tSZ Compton parameter panels is 1, for all other panels the summation of all curves is 0.\\
We observe that at large angular scales the recovery of tSZ $y$-map is driven by the lowest frequencies up to 255~GHz (shown in blue and green colors on Fig.~\ref{winunrel}).
For $\ell > 2000$ the reconstruction is dominated by the highest frequency channel that present the highest angular resolution.
For this small angular scale, we observe a general increase, by a factor of 5, of the weights for the linear combination.
This effect is particularly enhanced for the infra-red component cleaning (bottom panel).
Such an increase of the weights implies also an increase of the noise level and an increase in the leakage from other astrophysical components.

\begin{figure}[!th]
\begin{center}
\includegraphics[scale=0.35]{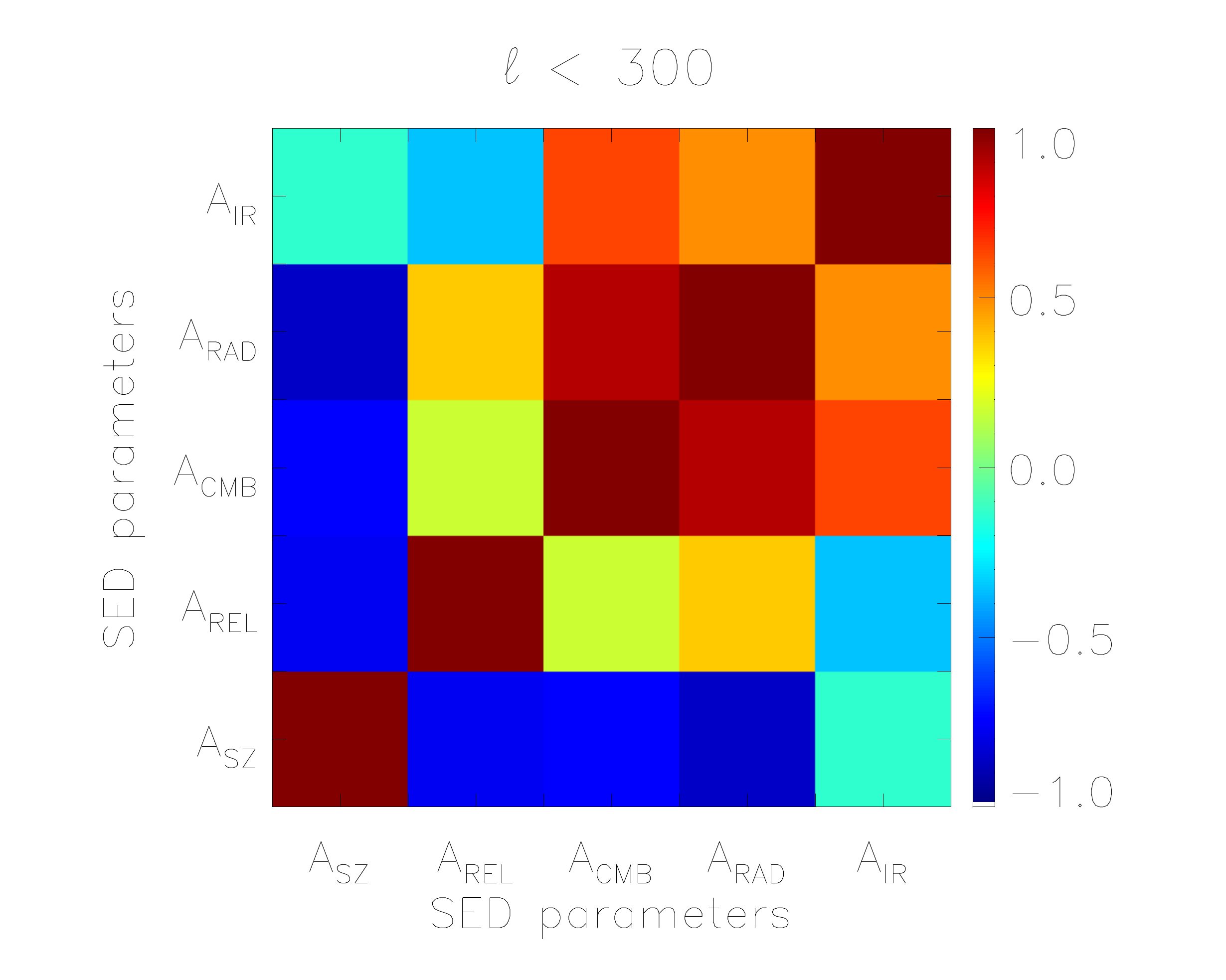}
\includegraphics[scale=0.35]{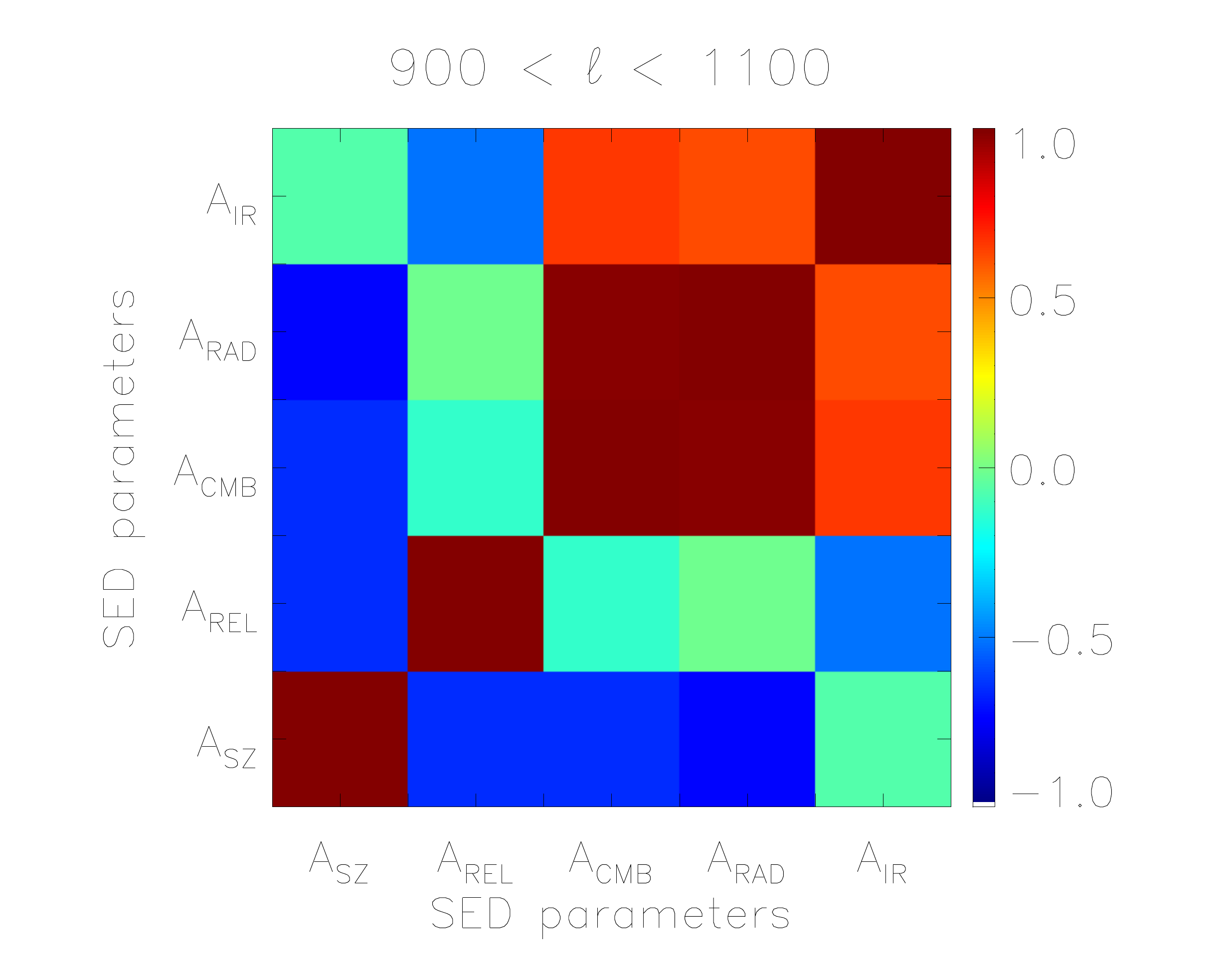}
\includegraphics[scale=0.35]{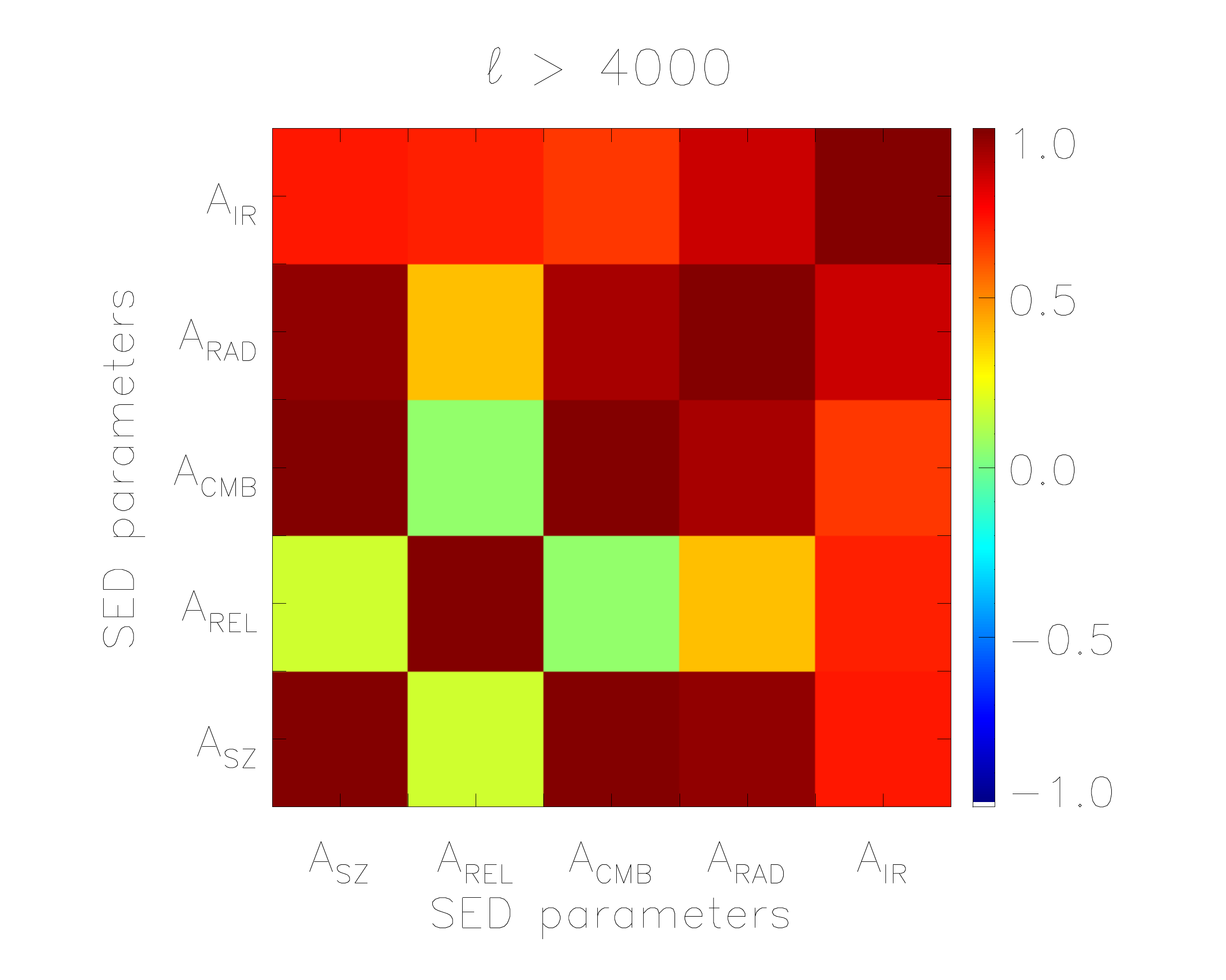}
\caption{From top to bottom: correlation matrices between the main astrophysical components for the reconstruction of tSZ $y$-map and $r$-map, at low-$\ell$ ($\ell < 300$), mid-$\ell$ ($\ell \simeq 1000$), and high-$\ell$ ($\ell > 4000$).}.
\label{cormats}
\end{center}
\end{figure}

\subsubsection{MILCA tSZ $r$-map}

Figure~\ref{winrel} presents the same information as Fig.~\ref{winunrel} for the tSZ relativistic corrections map.
Consequently, the sum of all the curves in the tSZ relativistic corrections panels average to 1, and to 0 for all other panels.\\
We observe that the tSZ relativistic corrections reconstruction is also dominated by the 255 GHz channel up to $\ell = 2000$. At larger angular scales, radio emission and CMB cleaning is performed using the lowest-frequency channels. At higher angular resolution, lowest frequency channels are not available anymore, and highest frequency are used, similarly as what we saw for the $y$-map, to reconstruct the tSZ $r$-map and especially clean for infra-red contamination. 

\subsubsection{Component noise covariance matrix}

The correlation in the noise of the recovered components is very informative on the potential leakage of a given component into another one. If we assume that all components are completely removed by the linear combination, the components noise covariance matrix, ${\cal V}_{\bf N}$, can be computed as
\begin{align}
{\cal V}_{\bf N} = \left( {\cal F}^{\rm T} {\cal C}^{-1}_{\bf N} {\cal F} \right)^{-1}.
\end{align}
This covariance matrix depends only on the astrophysical components SED properties, ${\cal F}$, and on the instrument frequency channels noise covariance matrix, ${\cal C}^{-1}_{\bf N}$.
Considering that the resolution of each observed channel varies, and assuming that the noise in each frequency band is not spatially correlated, ${\cal C}^{-1}_{\bf N}$ varies with the angular scale.
On Fig.~\ref{cormats}, we present the noise correlation matrices for: 
\begin{itemize}
\item tSZ Compton paramter, $A_{\rm SZ}$,
\item tSZ relativistic corrections, $A_{\rm REL}$,
\item CMB, $A_{\rm CMB}$,
\item a radio component, $A_{\rm RAD}$,
\item an infra-red component, $A_{\rm IR}$,
\end{itemize}
at 3 different angular scales: $\ell < 300$, $900 < \ell < 1100$, and $\ell > 4000$.
For low and mid-angular resolution correlation matrices, we observed a high level of anti-correlation between tSZ $y$-map and $r$-map. This is induced by the strong contribution of frequencies below 400 GHz where $y$-map and $r$-map have opposite sign for their contribution in the intensity maps. Similarly the $y$-map noise is anti-correlated with the CMB noise considering that the reconstruction is dominated by low-frequency channels. This behavior changes at high angular resolution, where we observed a significant level of correlation. The increase of the correlation level for high angular resolutions shows that the number of frequencies available for high angular resolution start to be too small and induces a more complexe separation of each component. 
In general, we observe that the reconstruction of the tSZ relativistic corrections are essentially limited by the non-relativistic tSZ effect and by the infra-red emission that mimics the high frequency part of the tSZ relativistic correction SED.

\section{Can we neglect tSZ relativistic corrections ?}

In this section, we determine the importance of considering tSZ relativistic corrections for Compton-parameter-focussed analyses. Are relativistic corrections a simple option or are they mandatory for the future of tSZ scientific analyses ?

\label{secbias}
\subsection{Component separation transfert function}
\label{relbias}
\begin{figure}[!th]
\begin{center}
\includegraphics[scale=0.15]{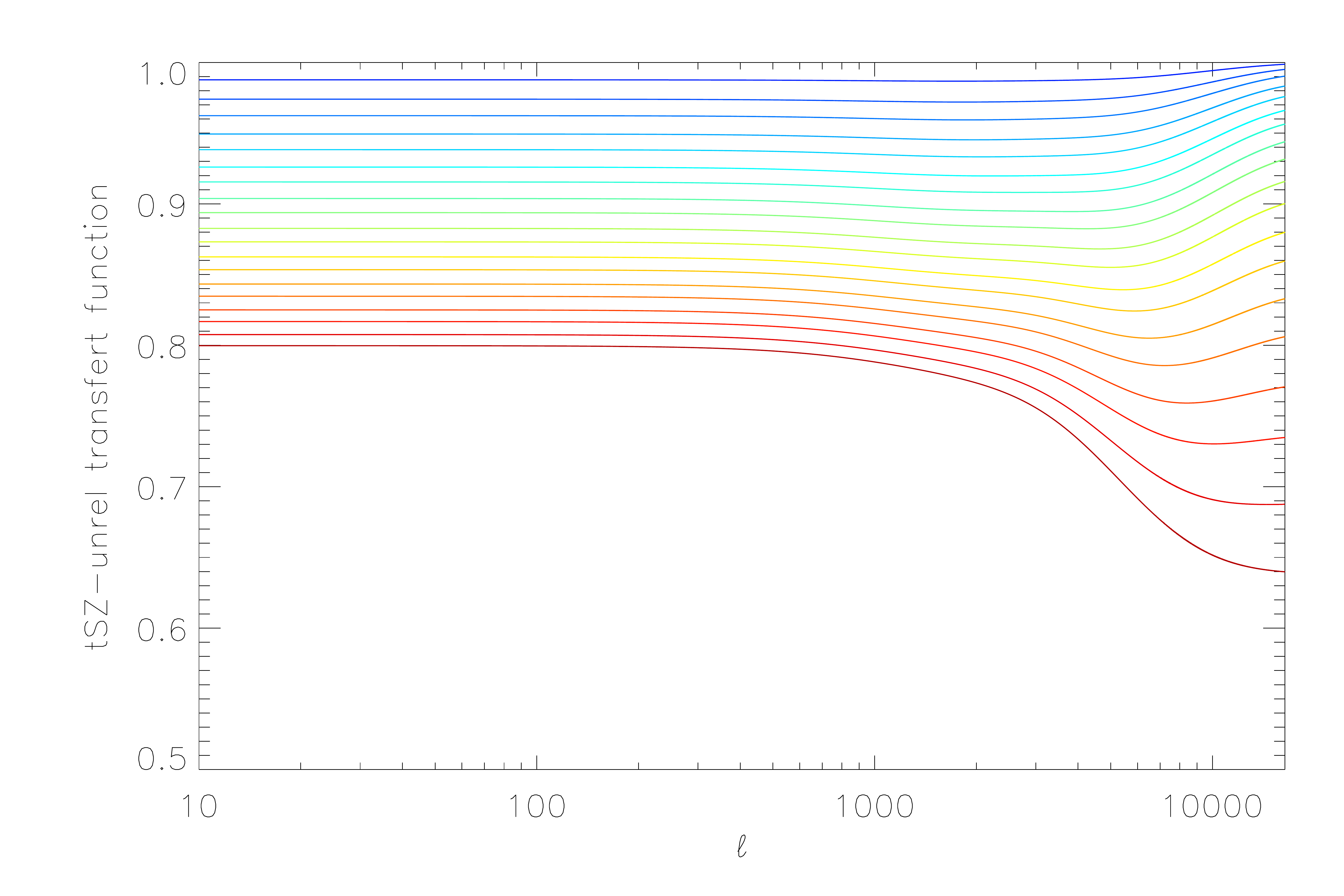}
\caption{Transfert function of the tSZ effect as a function of the scale $\ell$ and the temperature $T_e$, from 0 (dark blue) to 15 keV (dark red), when neglecting the relativistic during the reconstruction process for a COrE+.}.
\label{biasunrel}
\end{center}
\end{figure}

We applied Eq.~\ref{eqproj} to compute the transfert function of the real tSZ spectral distorsion, $g_{\nu, T_e}$ through a linear component separation that neglect relativistic corrections and only accounts for the non-relativistic tSZ spectral distorsion. 
Figure~\ref{biasunrel} shows the transfert function of the Compton parameter as a function of the temperature of the ICM and the considered angular scale.
For small temperature we observe a bias consistent with zero. However at high temperature we observe a significant bias of the tSZ flux of the cluster that reaches 20\% for $T_e = 15$ keV.
We also observe changes in the behavior of the bias at high-$\ell$ due the change of dominating frequencies in the reconstruction.
This bias only affects the overall normalisation and induces an under-estimations of the Compton parameter for high-temperature. The high angular-relation variation of the transfert function may also produce some distorsions of the tSZ profils in the core of galaxy clusters.
This under-estimation of the tSZ flux when neglecting tSZ relativistic correction can be easily understood from the SED of non-relativistic tSZ effect and tSZ relativistic corrections that reduces the amplitude of the tSZ spectral distorsion for frequencies below 400~GHz.
We stress that this bias is strongly dependent of the noise covariance matrix, as seen by the scale dependence of the bias at high-$\ell$ in Fig.~\ref{biasunrel}. Consequently, this bias is also strongly dependent of the setting used for the experiment.
For example, in the case of the Planck experiment, we derived that the induced bias is about a factor of two smaller than for a COrE+ like experiment.

We note that the galaxy clusters tSZ fluxes in the context of the Planck mission have been calibrated without considering the relativistic corrections, thus these tSZ fluxes are consistent with the fluxes in Planck tSZ $y$-map.
This has been shown in \citet{planck15szmap} over a large range of $\ell$ from 100 to 1000.
Consequently, cosmological constraints extracted using the $Y-M$ scaling relation calibrated on the Planck data \citep{planckszc,planck15szmap} are not affected by this bias, as all the bias cancels due to scaling relations being calibrated on the same data.\\
However, this challenges the meaning of using hydrodynamic-simulations \citep[see e.g.,][]{hor16} to constraint cosmological parameters. Indeed, the tSZ intensity in the Planck tSZ $y$-map, that neglects relativistic corrections \citep{planck15szmap}, is biased (up to 20\% at 15 keV) compared to the total amount of pressure in galaxy clusters.\\

\subsection{Scaling relation and power spectrum distorsion}

We have shown in Sect.~\ref{relbias} that tSZ fluxes are biased when the relativistic corrections to the tSZ spectral distorsion are neglected. Given that the amplitude of this bias is related to the temperature of the ICM, this bias also induces distorsions on the mass-observable relation, $Y-M$, and on the tSZ power spectrum shape.\\

When neglected, relativistic corrections to tSZ effect produce an under-estimation of high-temperature galaxy clusters Compton parameter that leads to a lower slope for the $Y-M$ relation.
For scaling relation, this bias depends on the considered mass range, redshift range, and experiment. For a COrE+ like experiment this bias would be $\simeq -0.05$ on the scaling relation slope, for galaxy cluster masses between $10^{14}$ and $10^{15} \, M_\odot$ and redshift below 0.5. For the Planck experiment, we estimated a bias of $\simeq -0.03$ on the $Y-M$ slope.\\

\begin{figure}[!th]
\begin{center}
\includegraphics[scale=0.15]{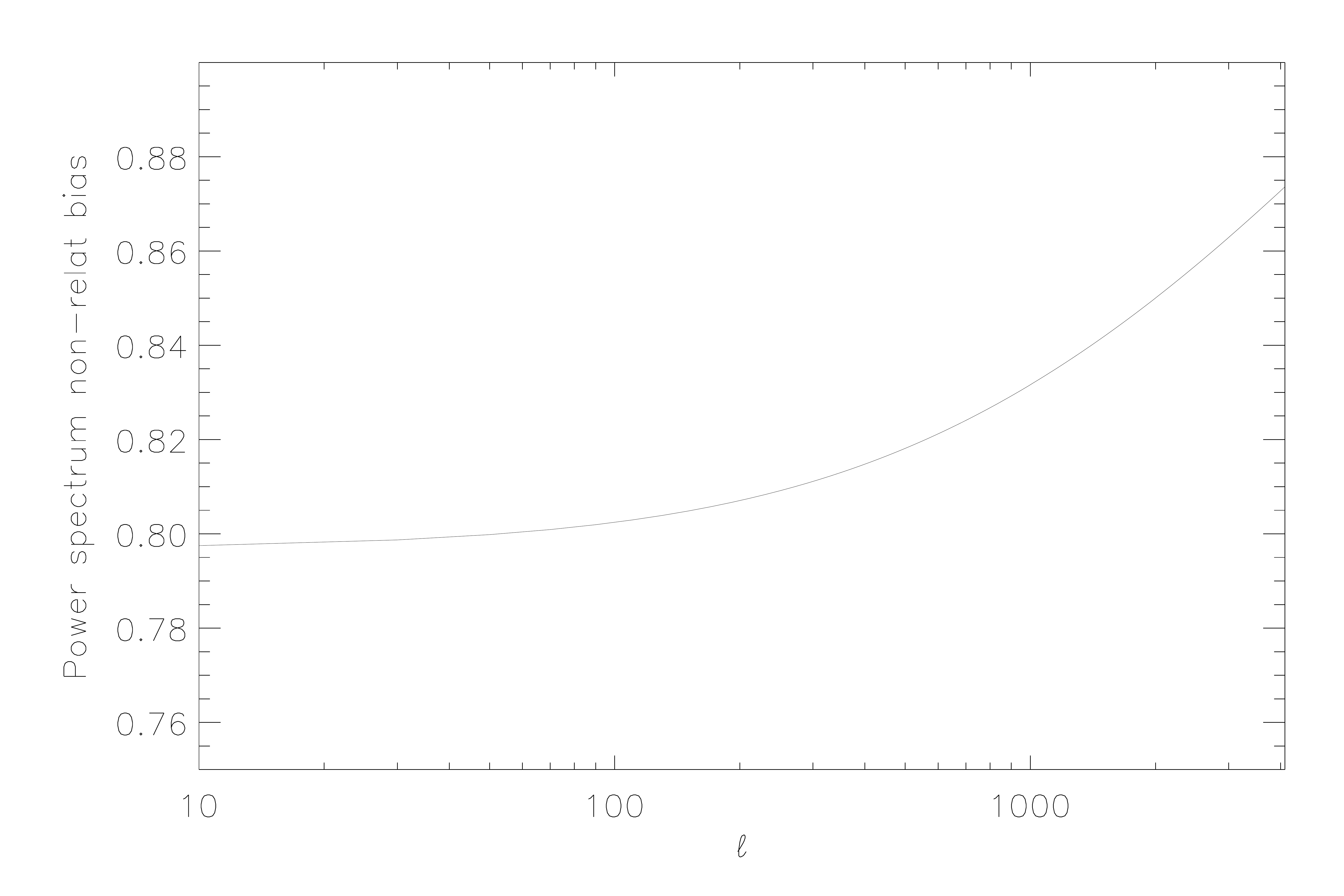}
\caption{tSZ auto-correlation angular power spectrum distorsion induced by relativistic corrections in component separated tSZ $y$-map.}
\label{clbias}
\end{center}
\end{figure}

Similarly to \citet{planck15szmap}~and~\citet{hur15}, we used a halo model to predict the tSZ angular power and characterize the impact of relativistic corrections on the tSZ angular power spectrum.
On Fig.~\ref{clbias}, we present the ratio between the model that account for the relativistic corrections induced bias, and the model neglecting relativistic corrections for the same $Y-M$ mass-observable relation. We observe that at low-$\ell$ the power spectrum is dominated by objects that present a 10\% bias (corresponding to $\simeq 20$\% bias on the tSZ power spectrum) on the Compton parameter flux when neglecting relativistic corrections.
We stress that the overall normalisation of this figure is not relevant due to $Y-M$ calibration process on the data. However, the distorsion is clearly observed, as a consequence of a modification of the slope of the $Y-M$ relation.
Consequently, accounting for relativistic corrections of the tSZ spectral distorsion is not optional for future experiments, but is required for precision astrophysics that can be done with the tSZ effect, and is mandatory to perform relevant comparisons with numerical simulations.

\section{Systematic effets on reconstructed relativistic corrections map}

\label{secsyst}
We have shown that MILCA allows to recover the tSZ signal and tSZ relativistic corrections at high-S/N ratio assuming the sensitivity of the COrE+ experiment, improving the sensitivity of Planck to the tSZ effect by a factor of 10.

\subsection{Non-linearity}

The main systematic effect comes from our linear approximation of the tSZ relativistic correction approximation.
Indeed, the amplitude of relativistic corrections are not exactly linearly related to the temperature of the ICM.
To estimate the component mixing and the effective transfert fonction of the $r$-map, we applied Eq.~\ref{eqproj} to compute the transfert function of the real tSZ spectral distorsion, $g_{\nu, T_e}$ to our reconstruction of the tSZ  $y$-map and $r$-map.
We note that the non-relativistic tSZ effect is a well known spectra that do not varies from cluster to cluster, thus the related transfert function is equal to $\simeq 1$ and there is no leakage into other component maps at the calibration incertitude level of the concerned experiment. Similar argument apply for the CMB component.

\begin{figure}[!th]
\begin{center}
\includegraphics[scale=0.15]{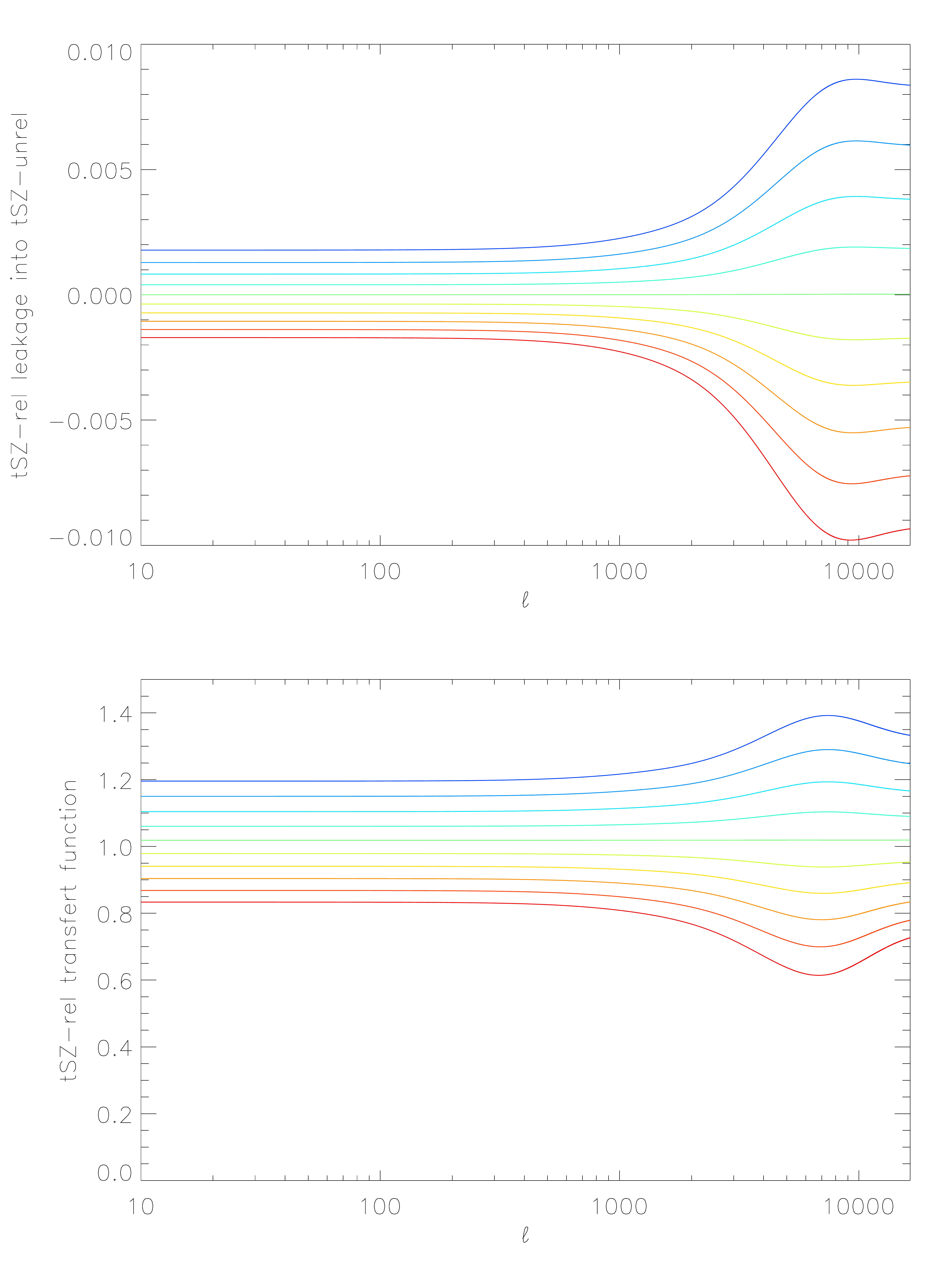}
\caption{Top panel: scale dependant tSZ relativistic corrections leakage into tSZ $y$-map. Bottom panel: scale dependant bias on tSZ relativistic corrections amplitude due to our first order linear approximation of the tSZ SED. Color ranging from blue to red indicates the considered temperature from 0 to 10 keV}.
\label{leakrel}
\end{center}
\end{figure}

On Fig.~\ref{leakrel}, we present the amplitude of the leakage of tSZ relativistic corrections into the non-relativistic tSZ $y$-map, and the transfert function of the tSZ relativistic corrections to tSZ $r$-map as a function of the ICM temperature.
We observe that the tSZ relativistic correction do not produce a significant bias into the tSZ $y$-map. The amplitude of relativistic correction is $yT_e$. Even at 10 keV, at low-$\ell$ only 0.2\% of the relativistic correction contaminates the $y$-map, this implies a bias of 2\% in the tSZ $y$-map. This leakage has to be compared with the 20\% of bias obtained when neglecting the relativistic corrections as shown by Figs.~\ref{biasunrel}~and~\ref{clbias}.
The leakage strongly increases at higher $\ell$, where the separation between non-relativistic and relativistic tSZ contributions is harder to achieve, leading to higher value for the weights of the linear combination, and thus to a higher level of bias in the case of SED mismatch.
We also observed that the transfert function for relativistic corrections amplitude in the $r$-map is almost flat up to $\ell = 2000$.
Thus this effect can be seen as a bias on the overall normalisation, the bias is monotonically related to the temperature, that allows to correct the galaxy clusters flux in the $r$-map, once the flux in the $y$-map is known. This is generally the case as the noise in the $y$-map is one order of magnitude below the noise in the $r$-map.
At first order, the bias can be corrected as follows
\begin{align}
\tilde{r} = (1+\alpha)\, r,  
\end{align}
with $\alpha$ an experiment dependent correction factor that account for the non-linearity of the tSZ relativistic corrections spectral distorsion with respect to the temperature of the ICM. In the present case, $\alpha \simeq 0.04 \,\left(\frac{r}{y}-T_1\right)$. This non-linear correction factor depends on $r/y$ ratio. However, considering that pixels with a significant signal for $r$ will by construction have a high signal-to-noise ratio measurement of $y$.
Consequently, the $r/y$ ratio non-linear correction is not a limitation for the scientific exploitation of the tSZ relativistic corrections mapping.

\subsection{Thermal dust emission}

\begin{figure}[!th]
\begin{center}
\includegraphics[scale=0.15]{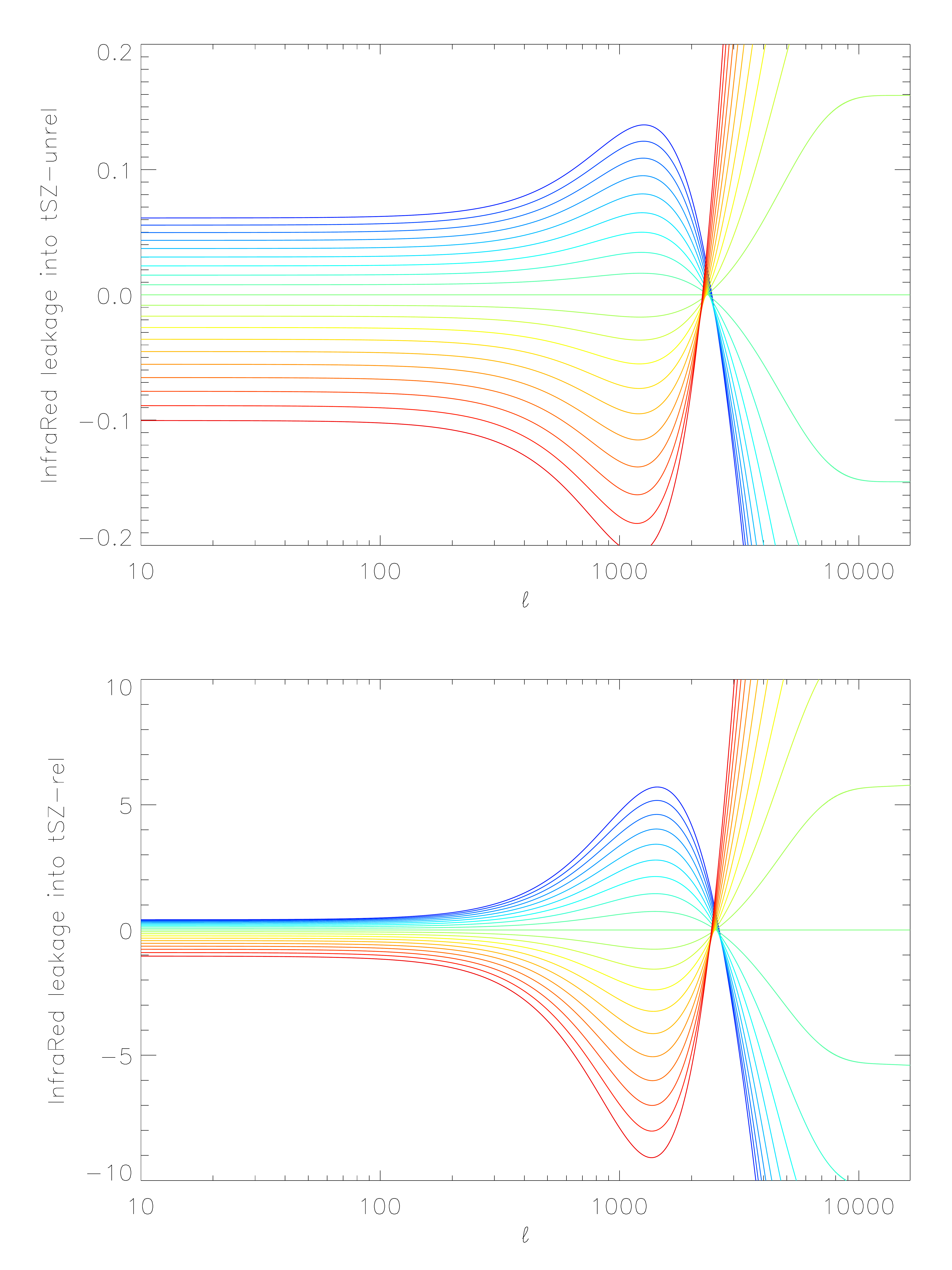}
\caption{Leakage of infra-red component into tSZ $y$-map (top panel) and $r$-map (bottom panel) as a function of the scale $\ell$ for a mis-estimation of infra-red emission spectral properties. Color ranging from blue to red indicates the considered error on the thermal dust emission grey body spectral index, $\beta_{\rm d}$, from -0.2 to 0.2. An intensity of $1\, K_{\rm CMB}$ at 195 GHz is used as a reference for the infra-red emission.}
\label{leakir}
\end{center}
\end{figure}

The contribution of the infra-red components in the MILCA $y$-map and $r$-map is crucial for tSZ relativistic corrections measurement. If the contributions from galactic thermal dust and high-$z$ CIB will naturally be reduced by the variance minimisation of the recovered maps, the contribution of the infra-red emission produced by galaxies inside the clusters \citep[see,][for a detailed analysis]{planckszcib} is more problematic, and is significant even for intermediate redshift galaxy clusters. 
By construction, this emission is spatially correlated with the tSZ effect. Such spatial correlation between astrophysical components leads to significant bias when applying blindly a minimum variance estimator, considering that the infra-red signal can be used to remove the tSZ signal and thus reduce the variance in the recovered maps.
Consequently, assumptions have to be made concerning the SED of the correlated thermal dust emission. These assumptions can be added in two ways in the MILCA method depending on the reliability of the thermal dust SED estimation: (i) hard constraints into the SED matrix, ${\cal F}$, or (ii) soft constraints by modifying the channel covariance matrix ${\cal C}_{\bf T}$ \citep[see][for more details]{hur13}.
For example, in a COrE+ like experiment the highest frequency channels from 600 to 800 GHz can be used to have an estimation of the thermal dust SED. However, this estimation will have uncertainties that will translate into bias in the tSZ $y$-map and $r$-map. We estimated this bias using Eq.~\ref{eqproj}. We assumed a modified black body SED for the thermal dust with a temperature $T_d = 20$ K and a spectral index $\beta_d = 1.7$. This values provides a fair description of the local univers infra-red emission from galaxy clusters \citep{planckszcib}.

We have seen that the weights of the linear combination are strongly varying with the considered angular scale.
Especially at small angular scale, the weights of the high-frequency becomes more important (due to their higher resolution).
Figure~\ref{leakir} presents the expected thermal dust leakage as a function of the angular scale and of the error on the assumed thermal dust emission grey body spectral index for a 1 $K_{\rm CMB}$ sources at 195 GHz. 
At such frequency, the amplitude of the IR emission toward galaxy clusters is $\simeq 0.01-0.02$ Jy \citep{planckszcib}.
This leads, for a beam of 7 arcmin FWHM to an intensity at the peak of $10^{-6}$ K$_{\rm CMB}$ for the IR emission.
Thus from Fig.~\ref{leakir} we can deduce that an error of 0.2 on $\beta_d$ would induce a bias with an amplitude of $10^{-7}$ in the $y$-map and $10^{-6}$ keV in the $r$-map.

Below $\ell = 2000$ the amplitude of the leakage is small both in the tSZ $y$-map and $r$-map, and can safely been neglected.
However, at higher-$\ell$, where only high frequency remains for the component separation the bias strongly increase (we precise that the curves at high-$\ell$ on Fig.~\ref{leakir} are going off-scale by one order of magnitude).
This challenges the possibility to recover high-$\ell$ part of the non-relativistic tSZ effect and tSZ relativistic correction at small angular scales as infra-red contamination will becomes a threat for COrE+ like experiment.
This statement also challenges small-mass galaxy cluster detection. Indeed, small mass galaxy cluster are the most affected by infra-red contamination, and require high angular resolution to be detected.
Indeed, the $I_{\rm IR}/Y$ ratio scales as $M^{-0.7}$, as the IR flux ,$I_{\rm IR}$, scale as $M$ \citep{sha11}, and $Y$ scales as $M^{5/3}$ \citep{planckscal,sif13,sal15}.
The same arguments apply to high-redshift galaxy clusters. This will probably limits the resolution at which tSZ effect can be extracted, $\simeq$ 4 arcmin FWHM.

Spatially varying thermal dust SED will have a similar effect on the reconstructed maps. Indeed, only one main thermal dust SED is cancelled by the linear combination. Consequently, if the thermal dust varies across the skies this will add a bias proportional to the mismatch between the main thermal SED and the contaminating SED. We stress that the localisation processus of MILCA (spatial and angular scale localisation) aim at minimizing such effect.

\subsection{Radio emission}

Numerous galaxy clusters host a bright radio-loud AGN at their center, well known nearby galaxy clusters in this situation are Perseus and Virgo. Similarly to the infra-red emission from member galaxies, the radio emission from AGN will induce bias in the galaxy cluster tSZ signal reconstruction. This issue as been addressed since the tSZ reconstruction for the Planck experiment \citep{hur13}. Contrary to the infra-red emission, significant radio emission from clusters is produced by a single galaxy in the cluster. Consequently the radio emission from galaxy clusters, at least for resolved ones, can be separated from the tSZ effect using morphological criteria to measure the radio source SED properties. We model the radio galaxie SED as a power law $\nu^{\alpha_r}$, with $\alpha_r = 0.5$ in intensity units \citep{planckagn}.

\begin{figure}[!th]
\begin{center}
\includegraphics[scale=0.15]{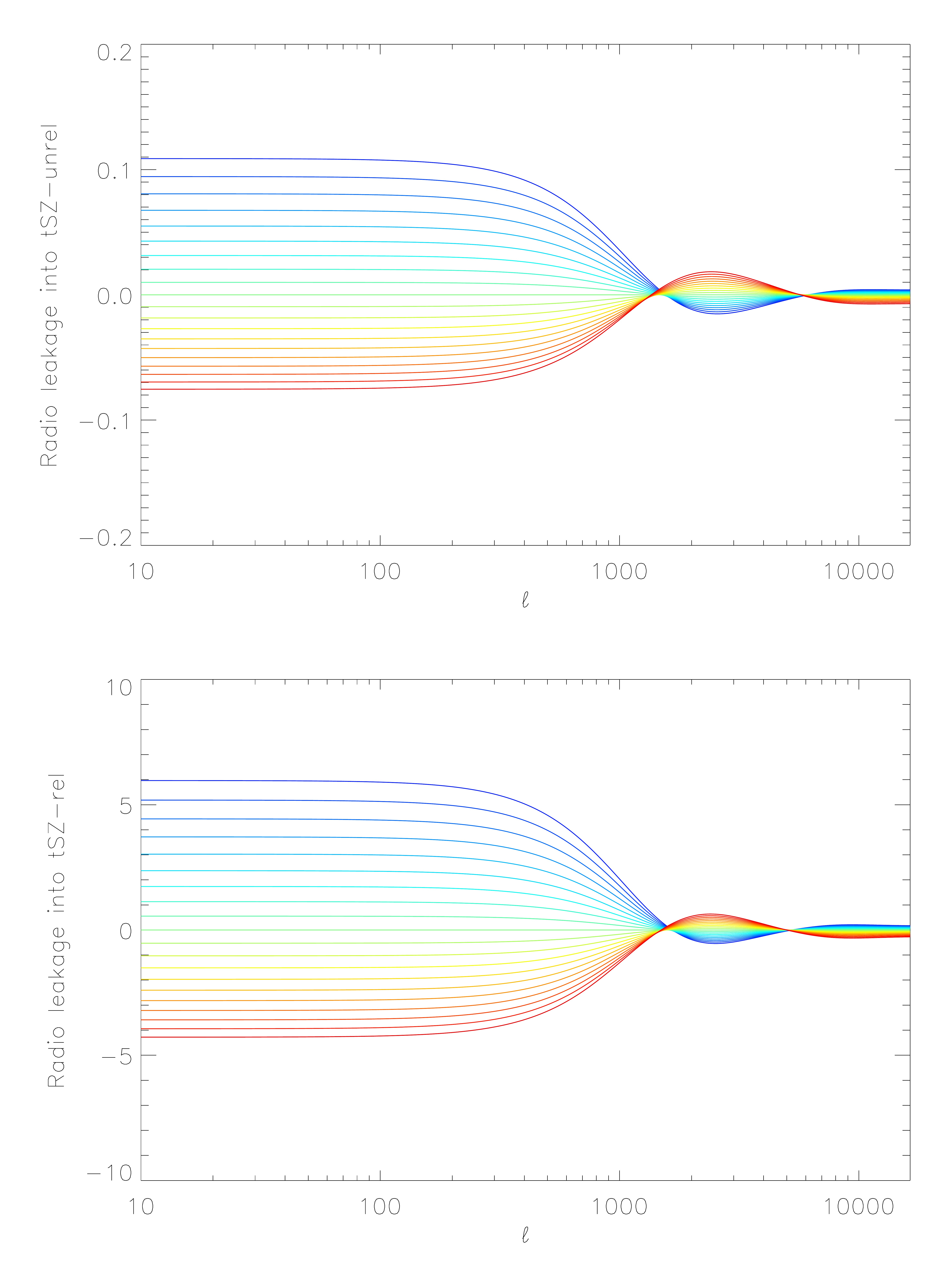}
\caption{Leakage of radio component into tSZ $y$-map (top panel) and $r$-map (bottom panel) as a function of the scale $\ell$ for a mis-estimation of radio emission spectral properties. Color ranging from blue to red indicates the considered error on the radio source SED spectral index, $\alpha_{\rm r}$, from -0.2 to 0.2. An intensity of $1\, K_{\rm CMB}$ at 195 GHz is used as a reference for the radio emission.}
\label{leakrad}
\end{center}
\end{figure}

The Figure~\ref{leakrad} presents the leakage of a radio sources into tSZ $y$-map and $r$-map as a function of the angular scale and the error on the spectral index of the radio source. We observe that the bias is higher at low-$\ell$ where the lowest-frequencies contributes the most to the reconstruction of tSZ maps.
For a bright radio source of 1 Jy at 1.4 GHz, the expected peak intensity with a 7 arcmin FWHM beam is $10^{-5}$ K$_{\rm CMB}$ at 195 GHz. This implies above $\ell = 1000$ for 0.2 uncertainties on $\alpha_{\rm r}$ the $y$-map will be contaminated up to $10^{-6}$ and the $r$-map up to $10^{-5}$ keV.
Similarly to the infra-red contamination we observe that the tSZ $y$-map present a small level of contamination that will at most correspond to $\simeq$1\% of the tSZ Compton parameter flux. The tSZ $r$-map is more significantly affected up to few percent below $\ell \simeq 1000$. Thus, tSZ relativistic corrections extraction will require a careful AGN contamination cleaning to avoid significant bias.\\

Combining the information from tSZ maps mixing, infra-red, and radio contamination we can identify that signal for $\ell$ below 2500 will enable a low contamination, low bias measurement of the tSZ relativistic correction for future experiments.

\section{Conclusion}

We have presented a new approach to extract tSZ-relativistic corrections using a component separation method, MILCA. This approach performed the reconstruction of the tSZ Compton parameter and relativistic corrections through a linear combination of multi-frequency datasets. This allows to provide robust estimations of the instrumental noise level and astrophysical component residuals (such as infra-red emission from member galaxies and radio-loud AGN emission from central galaxies).\\
We have characterized the MILCA transfert function to the tSZ relativistic corrections signal and performed a detailed analysis of the key frequency channels for both tSZ $y$-map and $r$-map reconstruction as a function of the angular scale for a COrE+ like experiment.\\
Our understanding of the foregrounds will be the key to measure at high accuracy and high angular resolution the relativistic correction to the tSZ spectral distorsion.
We identify that tSZ $y$-map and $r$-map can be accurately recovered with a low level of bias up to $\ell \simeq 3000$, for ICM temperatures ranging from 0 to 15 keV.\\

 Additionally, we have shown that relativistic corrections cannot be neglected for future high-sensitivity CMB experiments such as COrE+, as their impact on pressure measurement, tSZ scaling relation, and tSZ power spectrum will be significant even for Compton parameter focussed analyses. This bias is particularly important when comparison are made between tSZ data and numerical simulations.\\

We conclude that future high resolution experiments will offer a unique window on the tSZ effect spectral distorsion that will enables the possibility to perform detailed analysis of the ICM properties.
Such additional probes will be complementary with X-ray studies of galaxy cluster properties ({\color{blue} Tchernin \& Hurier in prep)}.
 
\label{concl}

\bibliographystyle{aa}
\bibliography{sz_relat_mapping.bib}

\end{document}